\documentclass[acmsmall,fleqn,screen]{acmart}

\usepackage{booktabs} 
\usepackage{url}
\usepackage{multirow}
\usepackage{siunitx}
\usepackage{amsmath}
\usepackage{enumerate}
\usepackage{algorithm}
\usepackage{algpseudocode}

\def\BibTeX{{\rm B\kern-.05em{\sc i\kern-.025em b}\kern-.08emT\kern-.1667em\lower.7ex\hbox{E}\kern-.125emX}}

\definecolor{rev}{rgb}{0.0, 0.0, 1.0}

\begin{document}

\title{Single-tap Latency Reduction with Single- or Double- tap Prediction}

\author{Naoto Nishida}
\affiliation{
  \institution{University of Tokyo}
  \city{Tokyo}
  \country{Japan}}
  \authornote{Both authors contributed equally to the paper.}
\email{nawta@g.ecc.u-tokyo.ac.jp}

\author{Kaori Ikematsu}
\authornotemark[1]
\affiliation{
  \institution{Yahoo Japan Corporation}
  \city{Tokyo}
  \country{Japan}}
\email{k-ikematsu@acm.org}

\author{Junichi Sato}
\affiliation{
  \institution{Yahoo Japan Corporation}
  \city{Tokyo}
  \country{Japan}}
\email{jsato@yahoo-corp.jp}

\author{Shota Yamanaka}
\affiliation{
  \institution{Yahoo Japan Corporation}
  \city{Tokyo}
  \country{Japan}}
\email{syamanak@yahoo-corp.jp}

\author{Kota Tsubouchi}
\affiliation{
  \institution{Yahoo Japan Corporation}
  \city{Tokyo}
  \country{Japan}}
\email{ktsubouc@yahoo-corp.jp}

\newcommand{\markupchi}[1]{\textcolor{black}{#1}}
\newcommand{\markupmobile}[1]{\textcolor{black}{#1}}
\newcommand{\markupcameraready}[1]{\textcolor{black}{#1}}
\newcommand{\technbig}{PREDICTAPS}
\renewcommand{\shortauthors}{Naoto Nishida et al.}
\newcommand{\yam}[1] {{\selectfont \color{red} #1 }}

\received{January 2023}
\received[revised]{May 2023}
\received[accepted]{June 2023}

\begin{abstract}
Touch surfaces are widely utilized for smartphones, tablet PCs, and laptops (touchpad), and single and double taps are the most basic and common operations on them.
The detection of single or double taps causes the {\it single-tap latency problem}, which creates a bottleneck in terms of the sensitivity of touch inputs. 
To reduce {\it the single-tap latency}, we propose a novel machine-learning-based tap prediction method called PredicTaps. 
Our method predicts whether a detected tap is a single tap or the first contact of a double tap without having to wait for the hundreds of milliseconds conventionally required. 
We present three evaluations and one user evaluation that demonstrate its broad applicability and usability for various tap situations on two form factors (touchpad and smartphone). 
The results showed PredicTaps reduces the {\it single-tap latency} from 150--500 ms to 12 ms on laptops and to 17.6 ms on smartphones without reducing usability.

\end{abstract}

\ccsdesc[300]{Human-centered computing~Interaction techniques}
\keywords{Latency Reduction; Single Tap; Double Tap; Prediction; Machine Learning; Single Tap Latency; Touch Surface.}

\setcopyright{acmlicensed}
\acmJournal{PACMHCI}
\acmYear{2023} \acmVolume{7} \acmNumber{MHCI} \acmArticle{224} \acmMonth{9} \acmPrice{15.00}\acmDOI{10.1145/3604271}

\maketitle

\begin{figure}[t]
\centering
\includegraphics[width=\columnwidth]{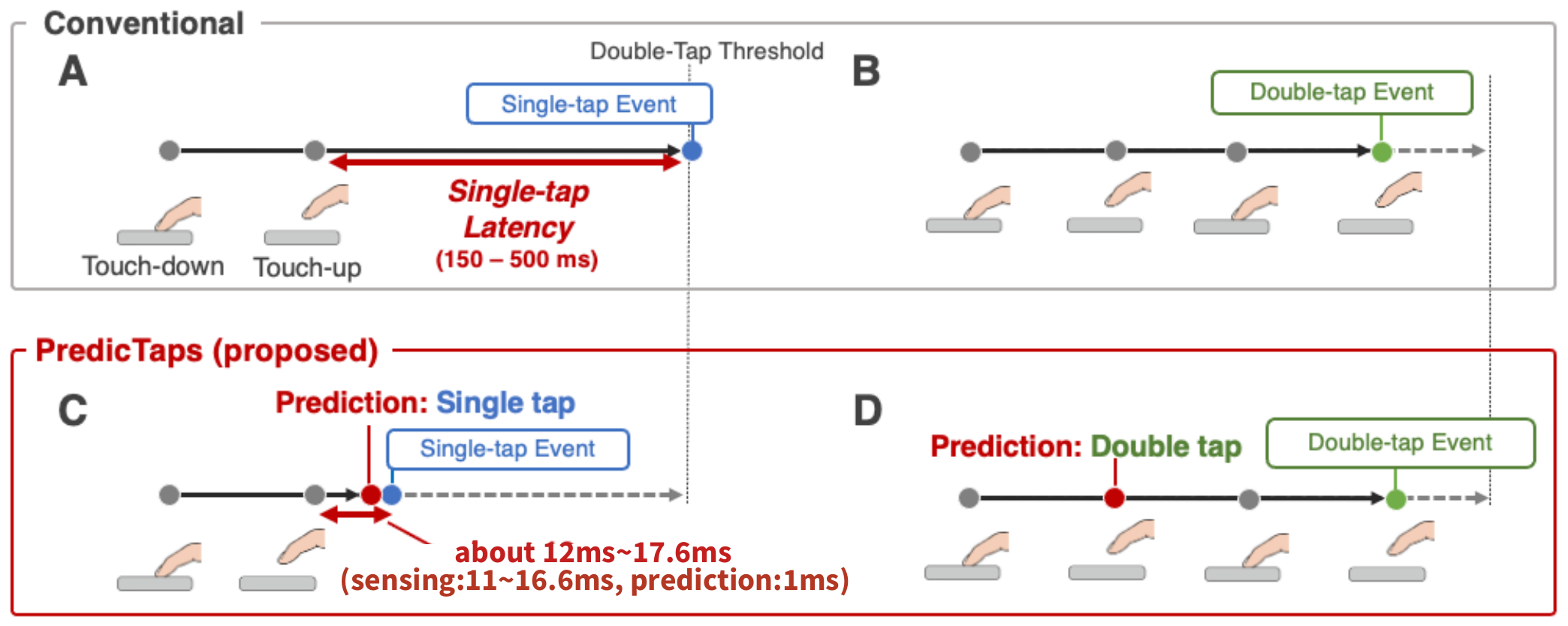}
\caption{
Processing with the conventional tap-detection method: (A) a single-tap event and (B) a double-tap event. Processing with the proposed method: (C) a single tap is predicted, so the system executes a single-tap event immediately after tap detection; and (D) a double tap is predicted, so the system waits for a subsequent tap.}
\label{feature-graphic}
\end{figure}

\section{Introduction}
Touch surfaces, such as touchpads, smartphones, and tablets, rely heavily on single and double tap inputs. 
Graphical user interface (GUI) apps on touch surfaces (e.g., e-book, map, painting, image viewers, and web browsers apps) generally have components that accept both tap types; single tap in a browser app is commonly used for selection, while double tap is used for zooming or displaying submenus.
Since touch surfaces typically have a limited space, To expand input options, multiple taps (i.e., double, triple, or more taps in a limited time) can be used, and touch prediction improvements are sought through hardware \cite{leigh14,TriboTouch} and OS \cite{tsoi2015advanced} research.

However, a problem known as the {\it single-tap latency problem} has hindered prediction speed improvements.  This can lead to bad usability when users require immediate responses (e.g., real-time strategy apps).
The conventional methods for distinguishing single and double taps can worsen this issue, as illustrated in Figure \ref{feature-graphic}.
In these methods, when the system detects a tap, it waits for a certain period of time for a subsequent tap to determine whether the detected tap was a single tap or the first tap of a double tap.
If a subsequent tap occurs within that period, the system recognizes the consecutive tap as a double tap, as seen in Fig. \ref{feature-graphic}(B); otherwise, it is recognized as a single tap (shown in (A)). 
With this type of distinguishing implementation, a double-tap event can be executed immediately after detecting the second tap (within 200--300 ms \cite{Heo:2014:ETI:2611205.2557234}), but a single-tap event requires a certain time period (e.g., 150--500 ms \cite{Heo:2014:ETI:2611205.2557234,MS,iOS}, usually predefined in the internal software configuration, hereinafter called the {\it double-tap threshold}).
Therefore, the time between the detection of a touch-up and the elapse of the {\it double-tap threshold} is considered the latency, i.e., {\it single-tap latency}.
Since single-tap operations are generally the most common, this latency is a critical problem: even such a small latency can lead to a significant time loss over a long period of time.

\begin{figure}[t!]
\centering
\includegraphics[width=0.4\columnwidth]{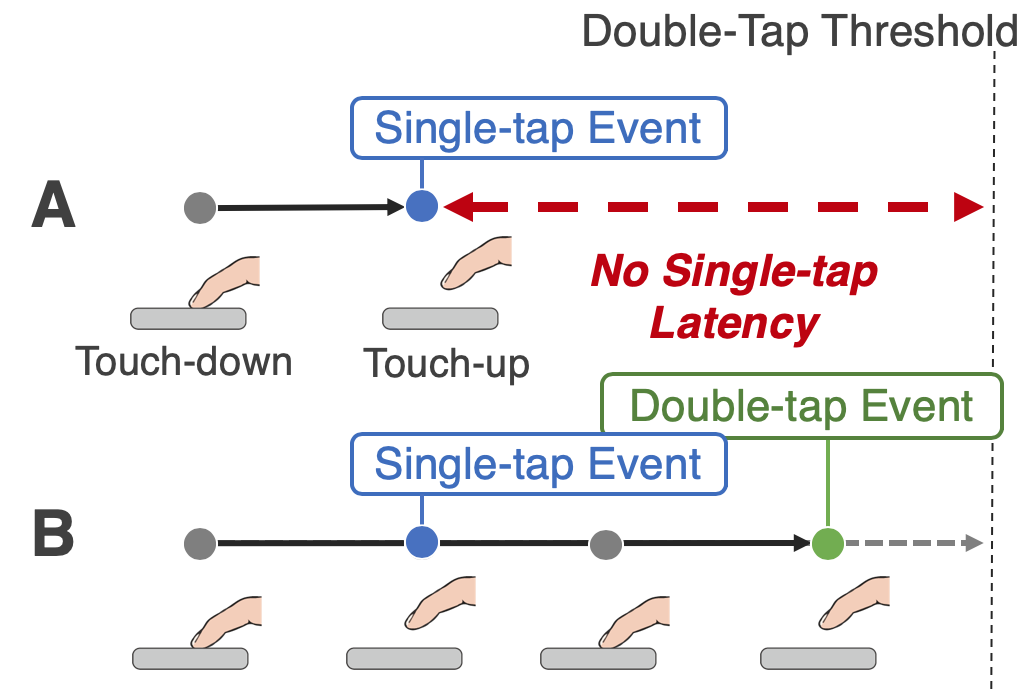}
\caption{
Processing method without {\it single-tap latency}. A single-tap event is executed immediately after every tap.
}
\label{fig:exception}
\end{figure}

Another problem is that consecutive single taps within a short time interval are sometimes misrecognized as double tap.
For example, when a user wants to move forward two pages in an e-book application, he or she has to carefully and slowly perform two consecutive single taps to avoid misprediction as a double tap.
Possible solutions include reducing the {\it double-tap threshold}, which may hinder double-tap execution \cite{smith1999aging} instead of speeding up the single-tap execution, or relying on UI designs and heuristics to bypass {\it single-tap latency}.
For example, the desktop and the Finder in macOS execute a single-tap event immediately after every tap, as shown in Figure \ref{fig:exception}, and iOS's WebView does not wait for double taps on certain web pages~\cite{iOS}. 
However, this processing is acceptable only when the single-tap event does not affect the double-tap event, and may not generalize well.

\begin{table}[]
\caption{
Experiments and corresponding papers.
}
\begin{tabular}{lll}
\hline
\textbf{Condition}                    & \textbf{Device}   & \textbf{Experiment}      \\ \hline
\multirow{2}{*}{ideal}       & laptop     & Ikematsu et al. \cite{Ikematsu:2020,Ikematsu:2020jp} \\ \cline{2-3} 
                             & smartphone & this paper      \\ \hline
\multirow{2}{*}{in the wild} & laptop     & this paper      \\ \cline{2-3} 
                             & smartphone & this paper      \\ \hline
\end{tabular}
\label{fig:paper_experiments}
\end{table}

This study introduces PredicTaps, a machine learning-based method to reduce {\it single-tap latency} by predicting tap types based on sensing data.
By analyzing sensing data, the system can immediately predict whether a detected tap is a single tap or the first tap of a double tap after the first tap ends.
Then, the system decides whether to execute a single-tap event immediately (Fig. \ref{feature-graphic}(C)) or to wait for a subsequent second tap (D).
In this way, PredicTaps makes single- and double-tap events more practical without complex heuristics or detailed design.
While this basic idea of PredicTaps was proposed by Ikematsu et al. in 2020 \cite{Ikematsu:2020,Ikematsu:2020jp}, they only reported the results of a feasibility study on PredicTaps in ideal conditions free from noises on a touchpad.
To show the usefulness, PredicTaps still need to be evaluated under real-world conditions (e.g., accuracy, processing time, and robustness for various situations and for different form factors).
Moreover, it also needs to be evaluated the negative effect of inaccurate tap classification.
To assess the impact on the performance and usability in actual use and to investigate the generalizability of PredicTaps in detail, we focus on evaluations with data from different conditions and form factors.
The results showed that PredicTaps worked well in all experiments, which demonstrates its good usability and wide applicability to devices with only touch-related sensors.
PredicTaps showed 100\% accuracy by training with data with top 10\% high confidence scores on touchpads, and with top 50\% on smartphones.
In addition, in a user evaluation on smartphones, The participants stated positive opinions on PredicTaps.
Thus, the main contributions of our paper are as follows:
\begin{itemize}
\item As a solution to {\it single-tap latency problem}, we extended the evaluation of PredicTaps as a solution to the {\it single-tap latency} problem (Table. \ref{fig:paper_experiments}). 
\item We conducted three data collection experiments on two form factors (laptop touchpad and smartphone) and developed models to evaluate the robustness of PredicTaps in a variety of situations and form factors.
\item We conducted a PredicTaps user evaluation on smartphones, confirming it's latency reduction and improved usability. 

\end{itemize}

\section{Related Work}
\subsection{Latency and Perception}
The end-to-end latency inherent to tapping is defined as the time from when a user touches a surface to when the display changes accordingly. This latency affects the perception and performance in both direct (e.g., smartphone) and indirect (e.g., touchpad) touch interaction. According to Ng et al. \cite{Ng:2012:DLD:2380116.2380174}, the latency mainly arises from three components: (1) the physical sensors that capture touch input, (2) the software that processes touch events and generates output for the display, and (3) the display itself. With commercial touch screens, this latency ranges from 50 to 200 ms and is perceptible to their users \cite{Ng:2012:DLD:2380116.2380174,Deber:2016:HTL:2858036.2858394}. 

Previous works have investigated end-to-end latency perception in different form factors (i.e., direct touch input on a smartphone or indirect touch input on a touchpad) and tasks (i.e., tapping or dragging) \cite{10.1007/978-3-319-20681-3_13,Deber:2015:MFF:2702123.2702300,Jota:2013:FFE:2470654.2481317,Ng:2012:DLD:2380116.2380174}. Regarding indirect touch interaction, Deber et al. showed that the noticeable difference (JND, the minimum time difference between a pair of stimuli that are detectable by a person and measurable for any perceptual stimuli) with a touchpad is 55 ms for dragging and 96 ms for tapping \cite{Deber:2015:MFF:2702123.2702300}. They found that a latency improvement as small as 8.3 ms is noticeable for a wide range of baseline latencies. They also showed that direct touch interaction is more affected by end-to-end latency, as a user can easily notice the physical distance between his or her finger and the system's output, particularly when dragging. The JND with a touchscreen in their study was 11 ms for dragging and 69 ms for tapping \cite{Deber:2015:MFF:2702123.2702300}. Jota et al. showed that performance is negatively affected when the latency is above 25 ms for dragging tasks \cite{Jota:2013:FFE:2470654.2481317}. 
Overall, all modern touch surfaces suffer \textit{single-tap latency}, meaning that latency reduction is a highly relevant goal. 

Furthermore, the end-to-end latency in tapping (single taps) on a double-tappable component increases because of the additional {\it single-tap latency}. 
In general, the {\it double-tap threshold} at the OS level can be configured by the user from the accessibility settings. Specifically, the default settings are 250 ms (in the range from 200 to 500 ms) for iOS~\cite{apple} and 500 ms (from 100 to 900 ms) for Windows~\cite{MS2}. 

In macOS, the default threshold is estimated at to be more than 150 ms, and it can be configured through the accessibility settings.
In contrast, an application-level threshold is often set separately, e.g., 350 ms for WebKit on iOS~\cite{iOS} and 300--500 ms for Safari~\cite{Heo:2014:ETI:2611205.2557234,ionescu2016using}. 
As with single-finger taps, there is a similar delay for multi-finger taps (e.g., Smart Zoom on macOS)\footnote{Smart Zoom (Apple Inc.), \url{https://support.apple.com/en-us/HT204895}}. 
In addition to the major OSs mentioned already, many other systems have introduced processing using the {\it double-tap threshold} ~\cite{COCKBURN201921,tappat}. 
Although the threshold varies depending on the OSs, applications, and form factors, in most cases it is a few hundred milliseconds. 
Since single taps are one of the most frequently used operations, even such a small latency can lead to significant time loss over a long period of time. 

\subsection{Latency Reduction for Touch Interactions}
Previous works have also investigated how to reduce end-to-end latency with touch surfaces. One approach utilizes hardware, such as combining high-speed cameras with high-speed projectors to visualize finger inputs \cite{Kusano:2015:TUI:2817721.2817724,Ng:2012:DLD:2380116.2380174}; achieving the latency of 1 ms. Another approach is based on predicting user actions, such as next-points of a finger, to reduce the latency. 
According to Nancel et al. \cite{Nancel:2016:NPM:2984511.2984590,Nancel:2018:NPD:3242587.3242646}, existing research on next-point prediction techniques can be classified into five main types: using neural networks \cite{Le:2017:PSR:3132272.3134138,Henze:2016:STL:2935334.2935381,Henze:2017:IST:3098279.3122150}, Taylor series \cite{Cattan:2015:RLC:2817721.2817736,touchpat4}, Kalman filtering \cite{Liang:1991:TRV:120782.120784,touchpat}, curve fitting \cite{touchpat2}, and heuristic approaches \cite{touchpat3}. 
These approaches have mainly focused on latency reduction for dragging tasks; for example, zero-latency tapping \cite{Xia:2014:ZTU:2642918.2647348} was intended to eliminate touch-down latency by using a 3D motion capture technique. 
In this paper, we focus on reducing the {\it single-tap latency} by applying machine learning to predict single or double taps without additional hardware. 

\subsection{Prediction-based Touch Interactions}
Researchers have proposed prediction-based techniques to augment touch interactions on touch surfaces. 
Various methods have been proposed: a posture estimation using IMU sensors \cite{10.1145/3242587.3242651}, and a vision-based hand shape recognitions \cite{10.1145/3132272.3134126}. 
Since the majority of modern touch surfaces utilize capacitive touch sensing, recent research has focused on the raw touch data from the touch driver (generally called capacitive raw image \cite{Guo:2015:CID:2817721.2817722,Holz:2015:BBU:2702123.2702518}). 
For example, various interaction techniques have been proposed using features of body parts: biometric authentication \cite{Guo:2015:CID:2817721.2817722,Holz:2015:BBU:2702123.2702518}, 
touch gestures \cite{Wang:2019:EFS:3290605.3300254}, 
detecting finger proximity \cite{Hinckley:2016:PSM:2858036.2858095}, 
detecting the force of a touch \cite{10.1145/3338286.3344389}, 
differentiating between finger and palm touches \cite{Le:2018:PUP:3173574.3173934}, 
identifying individual fingers \cite{Le:2019:IFF:3301275.3302295}, and 
estimating finger pitch and yaw \cite{Xiao:2015:EFA:2817721.2817737}. 
It is a promising means of capturing detailed touch event data such as finger contact size and shape.
However, accessing the capacitive raw image data requires kernel modification and is not widely available to developers.
In contrast, our technique uses only the touch events commonly provided by major OSs. 

\section{Methodology}
\subsection{Prediction for a Predetermined Action}

\begin{table}[b]
 \caption{Interplay of actual user behavior and system prediction.
 }
\begin{tabular}{lc|cc|}
\cline{3-4}
                                                                     & \multicolumn{1}{l|}{} & \multicolumn{2}{c|}{\textbf{Prediction}}                                                                                                                                               \\ \cline{3-4} 
                                                                     &                       & \multicolumn{1}{c|}{\textbf{Single tap}}                                                            & \textbf{Double tap}                                                              \\ \hline
\multicolumn{1}{|c|}{\multirow{2}{*}{\textbf{Actual user behavior}}} & \textbf{Single tap}   & \multicolumn{1}{c|}{\begin{tabular}[c]{@{}c@{}}(a) Single tap\\ (latency reduction)\end{tabular}}   & \begin{tabular}[c]{@{}c@{}}(b) Single tap \\ (same as conventional)\end{tabular} \\ \cline{2-4} 
\multicolumn{1}{|c|}{}                                               & \textbf{Double tap}   & \multicolumn{1}{c|}{\begin{tabular}[c]{@{}c@{}}(c) Single tap\\ (unintentional input)\end{tabular}} & \begin{tabular}[c]{@{}c@{}}(d) Double tap\\ (same as conventional)\end{tabular}  \\ \hline
\end{tabular}
\label{table:combination}
\end{table}

Performing a double tap involves tapping a surface twice within a certain period. In general, the minimum amount of time between a visual stimulus and movement is estimated to be 260--290 ms \cite{keele1968processing,schmidt1978sources}. 
Since a double tap must be completed within 200--300 ms \cite{Heo:2014:ETI:2611205.2557234}, a user wanting to perform a double tap has to consciously decide to make two consecutive taps before making the first tap. This means that the double-tap motion can be described as a {\it predetermined} action. 
A double tap is thus likely to be a faster motion than a single tap. Thus, the differences between single- and double-tap actions should influence the touch event-related data (e.g., tiny finger movements on the touchpad, touch-down to touch-up, or a finger's contact area). 
PredicTaps uses this touch event--related data to predict whether a detected tap is a single tap or the first tap of a double tap (see Section \ref{sec:data}).
This allows the apps to decide whether to execute a single tap event immediately or wait for a subsequent second tap, reducing {\it single-tap latency}.

The above is the basic processing of PredicTaps, but in the system design, we also need to consider the occurrence of erroneous prediction, since PredicTaps bases its processing on prediction by machine learning.
Table \ref{table:combination} lists the processing for each combination of the prediction and actual input in PredicTaps. 
In the case of a false-positive double tap (b), the processing is the same as in the conventional method with {\it single-tap latency}, so there is no degradation in terms of latency. 
In contrast, a false-positive single tap (c) causes an unintentional single-tap execution, leading to usability reduction. Therefore, we concluded that PredicTaps should be activated only when the prediction is highly probable. 

\subsection{Data Collection for PredicTaps}
\label{sec:data}

The PredicTaps system recorded the touch events while participants operated touch surfaces. 
The sampling rate was 90 Hz on a laptop touchpad and 60 Hz on an iPhone touchscreen due to the libraries. 
\markupmobile{We adopted the M5MultiTouchSupport\footnote{\url{https://github.com/mhuusko5/M5MultitouchSupport}} of macOS and the JavaScript Web API\footnote{\url{https://developer.mozilla.org/en-US/docs/Web/API}} to access finger inputs. The details of the collected data are listed in Table \ref{table:features}}. 

We assumed that performing a double tap is likely to be faster than a single tap because a double tap is a {\it predetermined} action. As such, we expected the completion time for a double tap to be shorter than that for a single tap, and ergo, that quick finger movements would lead to touching with a stronger force. 
We, therefore, expected that the maximum contact size (a value correlated to the major radius of an ellipsoidal contact point), the mean finger velocity from touch-down to touch-up (i.e., the distance between the touch-up and touch-down locations divided by the completion time), and the distance between the touch-down and touch-up locations (in percentages of the XY coordinates) would all be greater than those for a single tap. 
Moreover, we expected that the touch-down and up locations would probably differ between single tap and double tap, so we collected the touch locations at touch-down and touch-up (in percentages of the XY coordinates, with the touchpad's upper-left corner as the origin). 
In addition to the touch-event data, in the case of the laptop touchpad, PredicTaps recorded whether the laptop was connected to an AC adapter or running on battery power. This is because the touch sensitivity is slightly affected by the power source \cite{Ikematsu2019-zi,Grosse-Puppendahl:2017:FCG:3025453.3025808}. 

For the smartphone conditions, smartphone touchscreens feature direct input, and touch coordinates are affected by the position of the interactive elements on the screen. 
Therefore, we did not use touch coordinates in the model implementation.

In the training model phase, the PredicTaps system also collected the ground truth labels, i.e., data on whether a tap was single tap or double tap.
The threshold to distinguish a single tap from a double tap was set to 500 ms, as this is the time used in popular OSs for laptops \cite{MS}.

\begin{table}[]
\caption{Extracted features.}
\scriptsize
\begin{tabular}{llll}
\hline
\textbf{Form Factors}                & \textbf{Sensor Type}                  & \textbf{Sensing Frequency}      & \textbf{Features}                                                             \\ \hline
\multirow{7}{*}{Laptop}     & \multirow{6}{*}{Touchpad}    & \multirow{6}{*}{90 Hz} & Tapping completion time from touch-down to touch-up                  \\
                            &                              &                        & Maximum contact size                                                 \\
                            &                              &                        & Mean finger movement velocity from touch-down to touch-up (X, Y)     \\
                            &                              &                        & Distance between touch-down and touch-up locations (X, Y)            \\
                            &                              &                        & Touch location at touch-down (X, Y)                                  \\
                            &                              &                        & Touch location at touch-up (X, Y)                                    \\ \cline{2-4} 
                            & Battery status               & Dependent on event     & Connecting AC adapter or battery-power condition                     \\ \hline
\multirow{3}{*}{Smartphone} & \multirow{3}{*}{Touchscreen} & \multirow{3}{*}{60 Hz} & Tapping completion time from touch-down to touch-up                  \\
                            &                              &                        & Maximum contact size                                                 \\
                            \hline
\end{tabular}
\label{table:features}
\end{table}

\subsection{Learning Model for PredicTaps (Training Phase)}
\label{sec:learn}
For the machine--learning technique utilized to determine whether an initial tap is a single tap or the first tap of a double tap, we used 90\,\% of the data for training (10\,\% of which was for cross-validation) and the remaining 10\,\% for testing different classifiers. 
Tables \ref{table:features} list the features used for machine learning. 
For the non-time-series data, after standardization, we balanced the positive and negative samples in the training dataset, and then trained the models by using random sampling. 
Note that we balanced just the training data, not the number of records in the test dataset. 
To handle sampling randomness, we used the average value obtained from ten rounds of classification after the system had been trained. 
We optimized the cost parameter for logistic regression by 10-fold cross-validation.

To solve the classification problems, we used logistic regression with L1 regularization (LIBLINEAR v1.94). Note that our main aim here is not to propose a new learning method but to show how single and double taps can be detected with various sensor data; thus, we used an interpretable model, i.e., logistic regression.

\subsection{Score-Based Prediction from Learned PredicTaps Model  (Prediction Phase)}
\begin{figure}[t!]
\centering
\includegraphics[width=8cm]{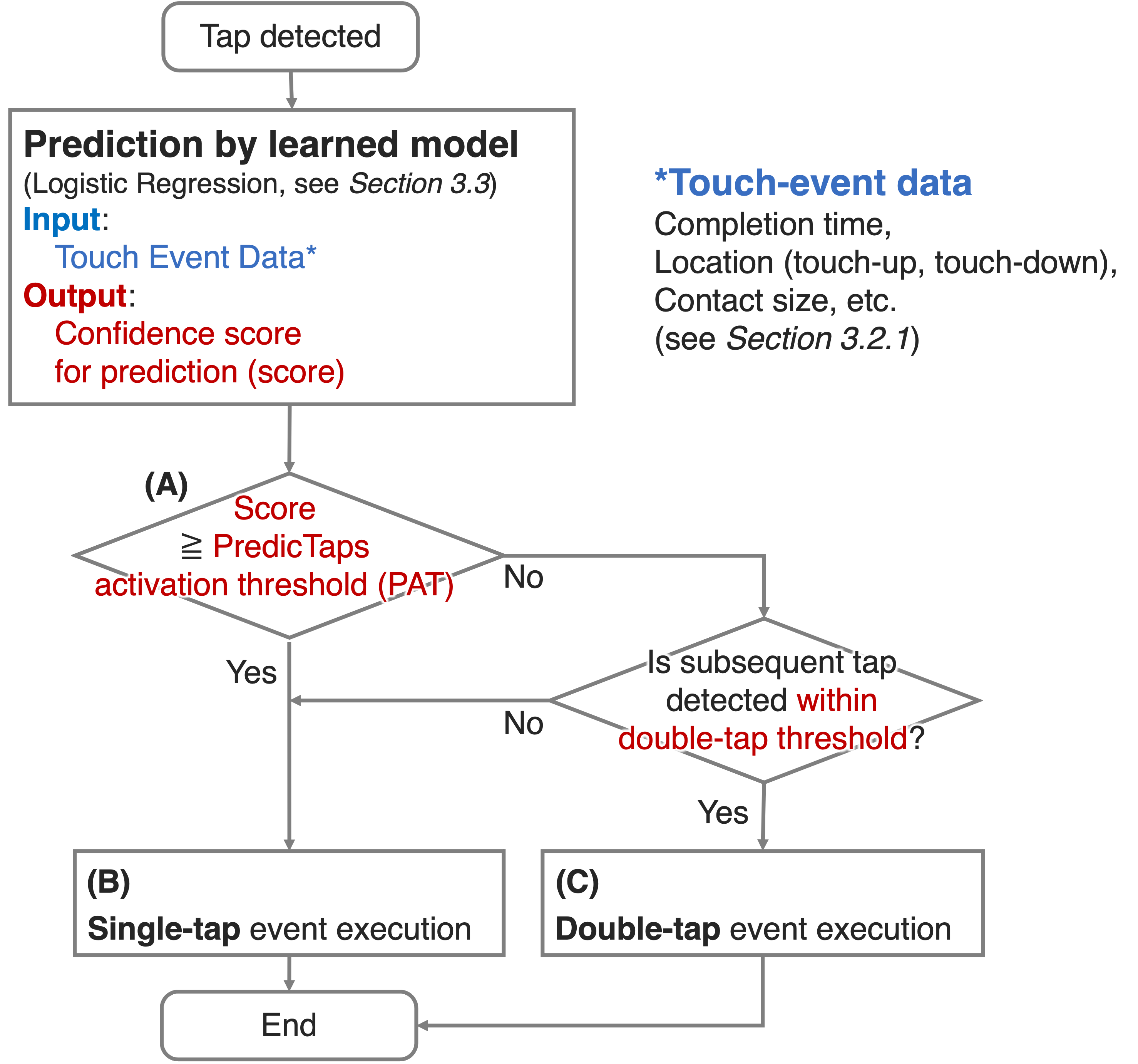}
\caption{
Process flow for determining whether a tap is a single tap or the first tap of a double tap using a trained model (in prediction phase). When the system detects a tap, it predicts whether the tap is a single tap or the first tap of a double- tap. If the system judges the tap to be a single tap with a confidence rate higher than {\it PAT}), a single-tap event is executed immediately without the {\it single-tap latency}.}
\label{fig:processing}
\end{figure}

In the prediction phase, PredicTaps instantly determines whether a tap is a single tap or the first tap of a double tap by retrieving the touch--event data and calculating a score. 
To reduce false-positive single taps, we set an additional threshold (called the {\it PredicTaps activation threshold} (PAT)) based on the confidence scores~\footnote{\url{https://scikit-learn.org/stable/modules/generated/sklearn.linear_model.LogisticRegression.html}}(hereafter \textit{score}). 
In logistic regression, this score is utilized for estimating the certainty that a tap is a single tap, and it is calculated according to the feature weights of a detected tap~\cite{king2008binary}. 
\markupchi{
In our system, if the calculated \textit{score} of a tap $ 0 \leq s \leq 1 $ is close to 1, it means a high possibility of a single tap, while an $ s $ close to 0 means a high possibility of a double tap. 
In normal logistic regression, when $ 0.5 \leq s \leq 1 $, the detected tap is recognized as a single tap. 
However, for good usability, the false--positive rate should be as low as possible, and the true--positive rate should be as high as possible. 
Defining the threshold as a hyperparameter, not a constant value 0.5, would result in models more tailored to individual users.
Therefore, we apply a threshold variable \textcolor{black}{ called the {\it PredicTaps activation threshold} (PAT)}. 
Then PredicTaps' prediction algorithm is calculated as follows:}
\markupchi{
\begin{equation}
  PredicTaps' prediction =
  \begin{cases}
    Single Tap & \text{if $ PAT \leq s \leq 1 $,} \\
    Double Tap & \text{if $0 \leq s \leq PAT $ } 
  \end{cases}
\end{equation}
}
A higher PAT reduces the number of false-positive single taps; that is, PredicTaps only accepts a tap as a single tap when the detected tap is highly likely to be a single tap (Fig.~\ref{fig:processing} (B)). 
In contrast, when the score is below the {\it PAT}---that is, when the reliability of the prediction is low, or when the detected tap is highly likely to be a double tap---the system waits for a subsequent second tap, the same as in the conventional approach (C). 

In PredicTaps, it triggers single-tap event only for taps with a high score, whereas it dismisses ambiguous taps or double-likely taps.
Although not all {\it single-tap latency} can be reduced by the above processing, the operability degradation due to misprediction of single and double taps can be prevented. We examined the effect of inconsistent latency reduction in a user study as well. In this evaluation, we examined PredicTaps accuracy using \textit{``data with the top n \% of the scores~($ 0 \leq n \leq 100$)''}. For example, the accuracy with data of the top 50 \% of the scores means that the accuracy is calculated using data with which PredicTaps is more confident than average about its decision. This accuracy was calculated as follows (see Appendix~\ref{sec:appendix} for the corresponding pseudocode). 
\markupchi{
\begin{enumerate}[Step 1:]
\item Calculate the absolute distance between $s$ and 0.5, as the $s$'s distance from 0.5 means greater confidence in the prediction of taps.
\item Sort the distance and the relevant data in descending order.
\item Extract data from one with the maximum distance until it fills n\% of the all data.
\end{enumerate}
}

In the following sections, we examine the robustness of PredicTaps in various conditions and form factors. We also evaluate the user experience of PredicTaps on smartphones.

\section{PredicTaps on Daily Laptop Use}
\label{sec:in-the-wild}
In the previous work by Ikematsu et al.~\cite{Ikematsu:2020}, they did not evaluate the performance of PredicTaps with data obtained in daily laptop uses; weakening the reliability of robustness of the system.
In order to evaluate the robustness of PredicTaps in daily use, we conducted an experiment to develop PredicTaps models and assess their accuracy on laptops.


\subsection{Participants and Apparatus}
We recruited \markupmobile{seventeen participants (twelve women, five men, sixteen right-handed, one left-handed, average age 25.16 with {\it SD} = 7.1). }
The participant used a MacBook Air, MacBook Pro, or MacBook model released in 2016, 2018, 2019, 2020, or 2021 with a screen size of either 13 or 14 inches, running macOS 10--13.
The laptops were equipped with an integrated touchpad and a logger app for data collection. 
We informed the participants that this experiment was to log the operation of the touchpad and obtained their consent to participate. 
\markupchi{Because touch sensitivity is slightly affected by the power source \cite{Ikematsu2019-zi, Grosse-Puppendahl:2017:FCG:3025453.3025808} and the computer was connected to the AC adapter at some times and was running on battery power at others, we, therefore, included the AC or battery-power condition in the features used for machine learning, with a value of 0 for AC and 1 for battery power. }

\subsection{Task Procedure and Data Collection}
To collect data on tapping during daily use of the touchpad, we developed a logging app that detects one-finger touch events. 
The participants installed the app and ran it while performing daily work on their laptops for four days.
The threshold to distinguish a single tap from a double tap was set to 500 ms. The participants were told not to use click and double-click actions and instead to use single- and double-tap actions. 
Even so, the participants were likely to sometimes click unintentionally, so we programmed the logging app to detect only tapping. 
We collected data for a total of 114,196 taps (single tap: 94,737; double tap: 19,459). 
None of the participants used triple taps for operations.

\begin{figure}[b!]
\centering
\includegraphics[width=\columnwidth]{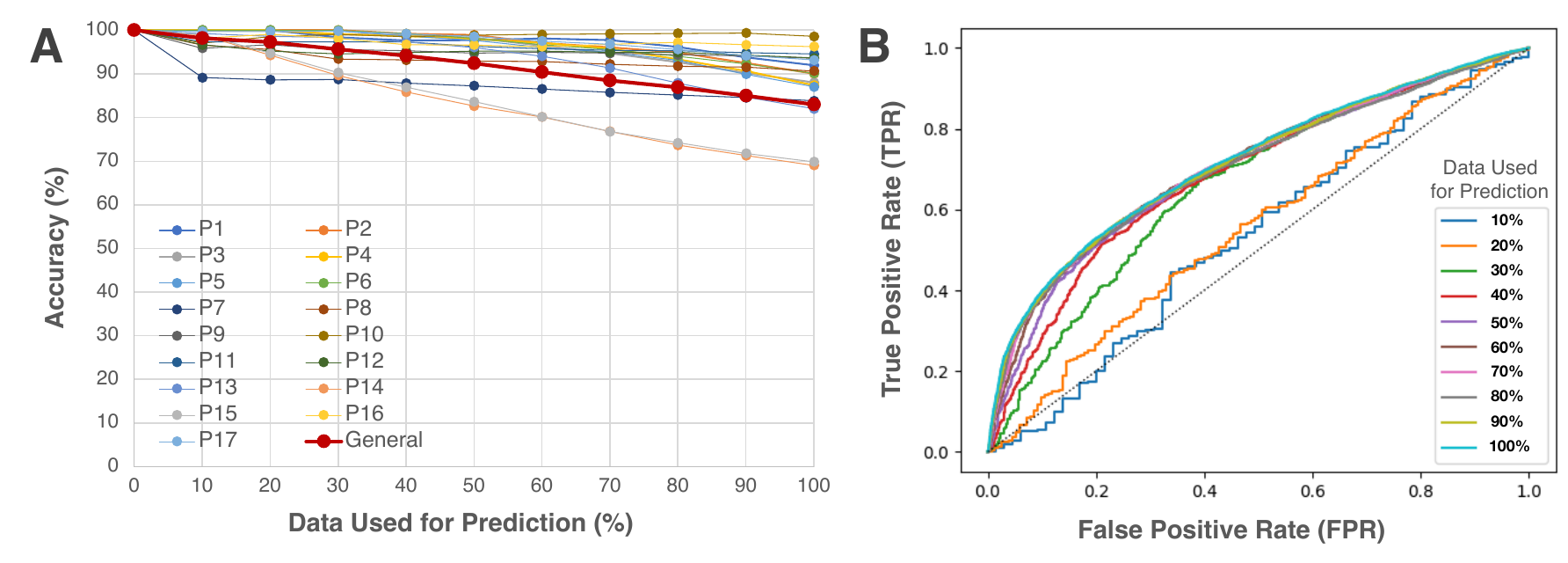}
\caption{
Accuracy and ROC curves for different percentages of data used for prediction in the uncontrolled experiment for a laptop touchpad. The red line in (A) is the accuracy of the general model trained with all participants' data, and the other lines are of the individual models. (B) is the ROC curve of the general model for different percentages of data used according to their calculated confidence scores (\it PAT). }
\label{models2}
\end{figure}

\subsection{Results}
We conducted the data processing in the same manner as discussed in Sec.~\ref{sec:learn}. 
Figure \ref{models2}(A) shows the prediction accuracy and (B) shows the receiver operating characteristic (ROC) curves for the developed model. 
In (A), the vertical axis represents the accuracy in predicting whether an initial tap was a single tap or the first tap of a double tap. The horizontal axis represents the amount of testing data obtaining high scores that were used for prediction. 
For example, the accuracy for a value of 10\% on the horizontal axis represents the prediction accuracy when only data in the top 10\% of scores were used. 
Note that, in logistic regression, the relationship between the accuracy and the score differs for each model. Although the accuracy (in \%) can be compared and discussed directly, the score itself cannot be compared between models. In addition, the distribution of the scores was not uniform; thus, we show how much accuracy could be obtained with respect to the data amount being used. 

The colored dots and red line in Fig. \ref{models2}(A) indicate the within-user accuracy and the general accuracy, respectively. 
The within-user accuracy means the prediction accuracy when the training and test data were from the same participant, and the general accuracy means the prediction accuracy when the training and test data were from all the participants. 
The {\it PAT} was applied to limit the data being used to high scores. 
As we can see, the accuracy gradually improved, reaching 95\% for the average of within-user cases when we limited the data to the top 50\%, and 95\% in the general case when we limited the data to the top 30\%.

Table \ref{table:feature_daily} lists the features used for machine learning along with their weights. 
The completion time was ranked as the top and bottom features for prediction. This means that an event with a longer completion time was more likely to be identified as a single tap.
Likewise, the maximum contact size was more likely to be identified as a double tap. This result matches our expectation that a double tap is a {\it predetermined} action. 
Besides, the touch--down location (X--axis) contributed to double-tap prediction. 
In contrast to the previous result, the velocity (XY-axis) contributed to single-tap prediction. 
Some features that were not selected by L1 regulation in the controlled experiment were used here for prediction.

It is possible that the differences in the above--mentioned features between our model and Ikematsu's model~\cite{Ikematsu:2020} were caused by the difference in tasks between them. Because the task in Ikematsu's experiment was a page-turning operation in an e-book reader, the users mainly performed single- or double-tap operations continuously and did not move the cursor frequently. In contrast, the participants here more frequently performed cursor movements and clicking. We presume that the differences between continuous tapping and point-to-tap operations could affect the tendencies of the features.

\begin{table}[t!]
 \begin{center}
 \caption{Weights for each feature in the experiment for laptop touchpad. }
 \small
 \begin{tabular}{c l   l} \hline 
\textbf{No.} &\textbf{Feature}&  \textbf{Weight}\\ \hline 
1& Tapping completion time from touch-down to touch-up &	5.615	    \\
2& Maximum contact size & --0.1186		    \\
3& Mean finger movement velocity from touch-down to touch-up (X, Y)&	(0.3420, 0.1698)	    \\
4& Distance between touch-down and touch-up location (X, Y)& (2.156,--0.2091)		    \\
5& Touch location at touch-down	(X, Y)&	(--1.477, 0.3351)	    \\
6& Touch location at touch-up	(X, Y)&	(0.8609, 0.2127)	    \\
7& Connecting AC adapter or battery-power condition &	0.000	    \\ 
\hline
\end{tabular}
\label{table:feature_daily}
\end{center} 
\end{table}

\section{Usability of PredicTaps on Smartphones}
\label{sec:smartphone}

In the previous work by Ikematsu et al.~\cite{Ikematsu:2020}, they only researched laptops.
To evaluate the robustness of PredicTaps by form factors, we conducted two experiments to assess PredicTaps user experience and smartphone performance.
In this section, we conducted a user study to examine how PredicTaps on smartphones affect usability because users easily notice the latency on direct touch screens, causing bad usability~\cite{Deber:2015:MFF:2702123.2702300, Jota:2013:FFE:2470654.2481317}.
We made two task applications for two distinct occasions: an \textit{Annotation Task} for when users need to think before tapping, and a \textit{Pointing Task} for when users can tap intuitively without thinking.
We collected training data and validation data on two days separately to evaluate if the system is robust under different conditions, and also conducted a user study on the second day.

\subsection{Participants and Apparatus}
Through SNS recruitment, we recruited 17 participants (eight women, nine men, 14 right-handed, two left-handed, one ambidextrous, average age 30.05 with {\it SD} = 8.069). We recruited participants from a broad range of ages and sex to ensure the system's generalizability and investigate the age-specific or sex-specific features. The average hours per day the participants used their smartphones was 4.41 ({\it SD} = 3.20). The average number of years the participants had owned their own smartphones was 9.118 ({\it SD} = 3.833). 16 participants used iPhone 13 (iOS 15.6.1) and one participant used iPhone 11 Pro (iOS 15.6.1) for this experiment.

\begin{figure}[b!]
\centering
\includegraphics[width=0.8\columnwidth]{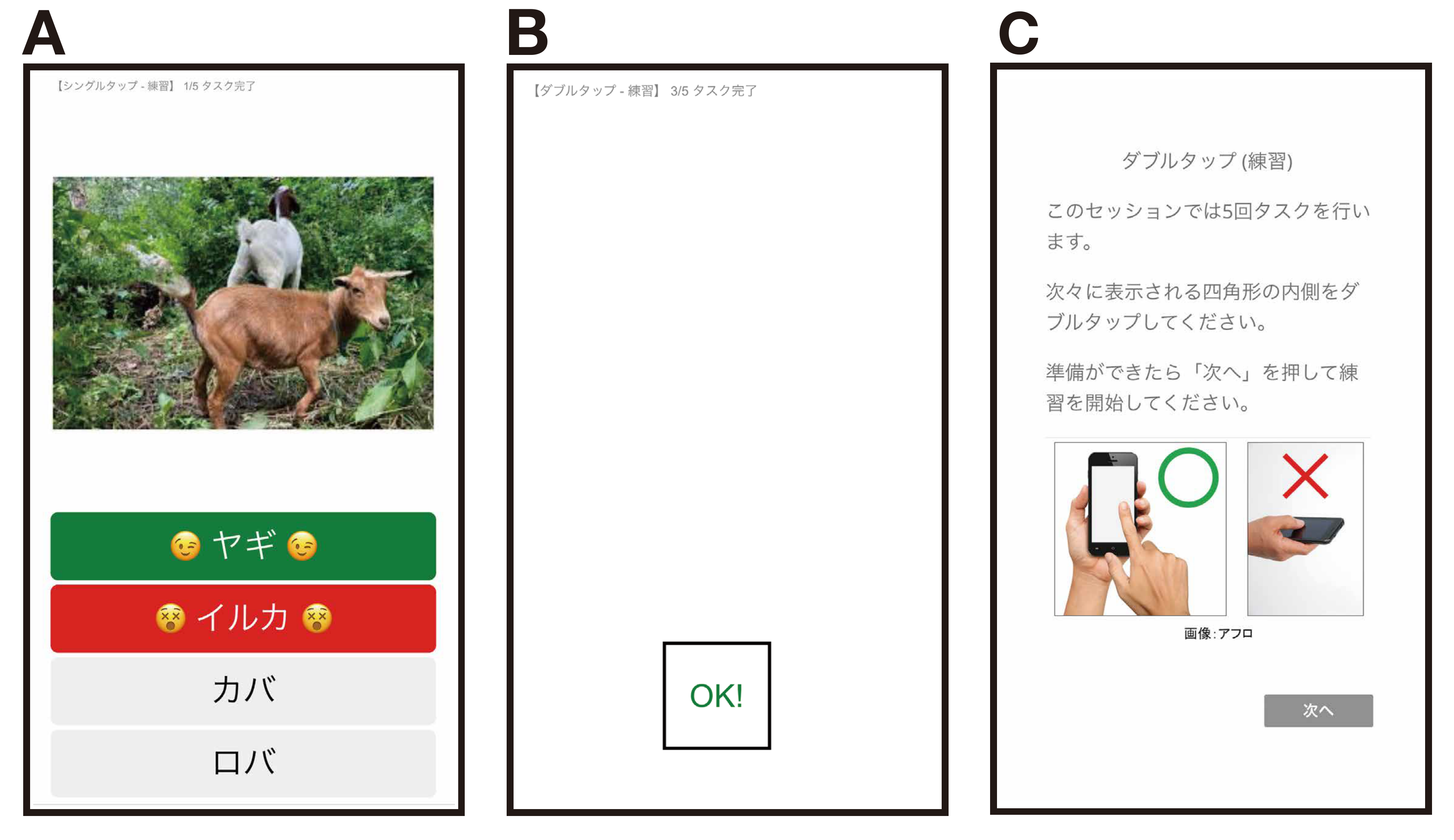}
\caption{
Screenshots of our experiment application. A) \textit{Annotation Task} part. The options are goat, dolphin, hippopotamus, and donkey.
B) \textit{Pointing Task} part. C) Instruction part. The instruction says, ``In this session, you need to tap five times. Double-tap the rectangles on the screen. Press the `next' button if you are ready." }
\label{ui_mobile}
\end{figure}

\subsection{Task Procedure}
\subsubsection{Day\,1: Training Data Collection}
\label{day1_smartphone}
On the first day, we collected training data of the tap input via two game apps: \textit{Annotation Task} and \textit{Pointing Task}. 
In the \textit{Annotation Task}, participants are required to pick the correct option according to the picture on the screen (Figure~\ref{ui_mobile}(A)). 
In the \textit{Pointing Task}, participants tap the rectangles (size: 1.6 cm $\times$ 1.6 cm) on the screen as quickly as possible (Fig. \ref{ui_mobile}(B)). 
They choose the answer or tap the rectangle by performing a single or double tap according to the instructions. 
There are 400 questions and rectangles in both tasks (200 single taps, 200 double taps for each task). 
The task order of tasks performed is randomized, and the order of single- or double-tap specification is randomized. 
In the instructions, the participants are told to use the index finger of their dominant hand to tap the options and rectangles, as well as to hold the smartphone in their non-dominant hand (Fig. \ref{ui_mobile}(C)). 
We also instructed them to hold and tap the smartphone as they would normally do but to try not to change their posture throughout the task.
After the instructions, each participant first performs one of the two tasks. 
The task is composed of a practice session followed by an experiment session. 
In the practice session, participants practice for ten taps (five single and five double) and then continue to perform in the experiment session, which consists of 400 taps (200 single and 200 single).
The participants take a 5-minute break and then perform the other task.
In each task, the participants need to stay on the same question or the same rectangle scene until they choose the correct answer or tap the rectangle correctly. 
The experiment lasted around 40 minutes for each participant. 
The average time for each task was 23.15 minutes ({\it SD} = 4.453) on the \textit{Annotation Task} and 14.52 minutes ({\it SD} = 5.735) on the \textit{Pointing Task}. 
We collected data of 13,600 taps in total (200 single taps and 200 double taps for one task; and one participant conducted two tasks; and 17 participants took part in this experiment).

\begin{figure}
\centering
\includegraphics[width=\columnwidth]{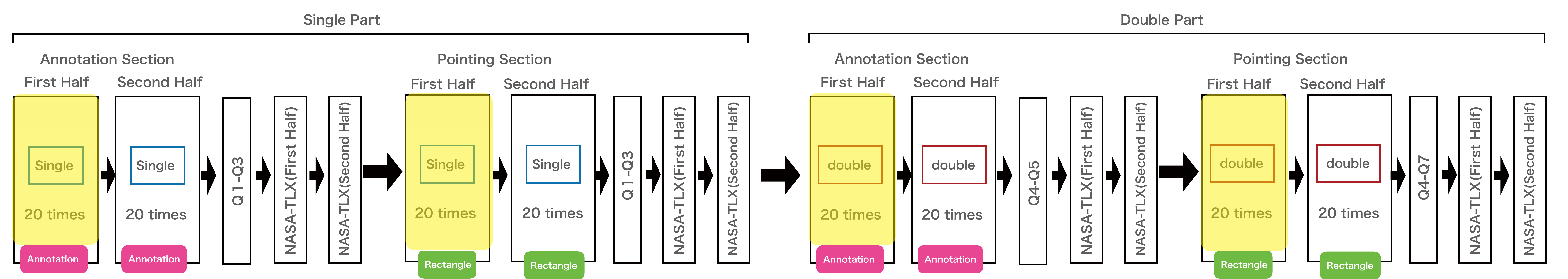}
\caption{
Workflow of the user study. It consists of two parts (Single and Double) and two sections (Annotations and Pointing). Each section is divided into a first and second half, and PredicTaps is activated in one or the other. The Single Part and Double Part/ Annotations Section and Pointing Section/ PredicTaps activation patterns are randomized. The yellow illustration is one example of the activation pattern.}
\label{userstudy_workflow}
\end{figure}

\begin{table}[]
\footnotesize
\caption{Questionnaire items for the user study. In Q2, Q3, Q5, Q6, and Q7, we used 7-point Likert scales.}
\begin{tabular}{lll}
\hline
   & \multicolumn{1}{c}{\textbf{Question}}                                                                                                                       & \multicolumn{1}{c}{\textbf{Reply Format}}                             \\ \hline
Q1 & Which partial section did you think PredicTaps was activated in?                                                                                      & The first half/The second half                               \\ \hline
Q2 & Rate your confidence level of the answer to Q1.                                                                                                    & 1:Fully unconfident -- 7:Fully confident   \\ \hline
Q3 & To what extent do you think the single-tap latency decreased?                                                                                      & 1:Not at all -- 7:Significantly decreases      \\ \hline
Q4 & Were your double taps incorrectly recognized by the system?                                                                                           & Yes/No                                                       \\ \hline
Q5 & (If ``Yes'' to Q4) To what extent can you accept the misjudgment?                                                                                   & 1:Fully unacceptable -- 7:Fully acceptable \\ \hline
Q6 & Rate the usefulness of PredicTaps.                                                                                                                 & 1:Worst - 7:Best                               \\ \hline
Q7 & \begin{tabular}[c]{@{}l@{}}When single tapping, did you feel the difference \\ between when PredicTaps activated and when it did not?\end{tabular} & Yes/No                                                       \\ \hline
Q8 & (If ``Yes'' to Q7) Did you get confused about the inconsistency?                                                                            & 1:Totally disagree -- 7:Totally agree          \\ \hline
\end{tabular}
\label{table:questionnaire_smartphone}
\end{table}

\markupchi{
\subsubsection{Day\,2: Validation Data Collection and User Study}
\label{day2_smartphone}
On the second day, we conducted a user study to examine if PredicTaps on smartphones affects the users' impressions or usability via two game apps that participants had already played on Day\,1. 
We implemented PredicTaps for each individuals, whose {\it PAT} wase set ranged from 0.6 to 0.7 depending on the occurance of false-positives of test data from Day\,1.
Half of the subjects participated approximately one month after the training data collection and the other half participated approximately one week later. The process of the experiment is illustrated in Fig. \ref{userstudy_workflow} and the questionnaire items for the user study are shown in Table~\ref{table:questionnaire_smartphone}. We segmented every pattern into four task groups: Single Tap / Annotation Task, Single Tap / Pointing Task, Double Tap / Annotation Task, and Double Tap / Pointing Task. The order of the task groups and whether the PredicTaps activation was in the first or second half of a section were randomized. We instructed participants to hold, tap, and maintain their posture, the same as in Sec. \ref{day1_smartphone}. We also instructed them to tap without a break in the specified task for 40 times per section. For example, in the Single Tap / Annotation Task group, participants performed the Annotation Task with single-tap only and did not stop until they finished 40 taps. Before the experiment, we explained how a touch surface detects a tap as either a single or double tap and briefly went over the PredicTaps mechanism. Specifically, we made sure they understood the following four points. 
\begin{enumerate}
    \item  PredicTaps predicts if the user's tap is likely to be a single tap or the first tap of a double tap.
    \item  PredicTaps can sometimes reduce the single-tap latency.
    \item  PredicTaps may sometimes mistakenly judge a double tap to be a single tap. 
    \item  PredicTaps is activated in either the first or second half of a section. 
\end{enumerate} 
After performing a task, 
participants were instructed to answer Q1 -- Q3 (in {\it Single Tap} groups) or Q4 -- Q5 (in {\it Double Tap} groups) questions and NASA--TLX for the first and the second halves, respectively. We processed this in the other task groups repetitively. The average total time for the experiment was 52 minutes (\textit{SD} = 4.6).
After the participants completed four task groups, the participants are instructed to answer Q6 -- Q8.
We collected data of 2,720 taps in total (80 single taps and 80 double taps; and 17 participants took part in this experiment).}


\begin{figure}[b!]
\centering
\includegraphics[width=\columnwidth]{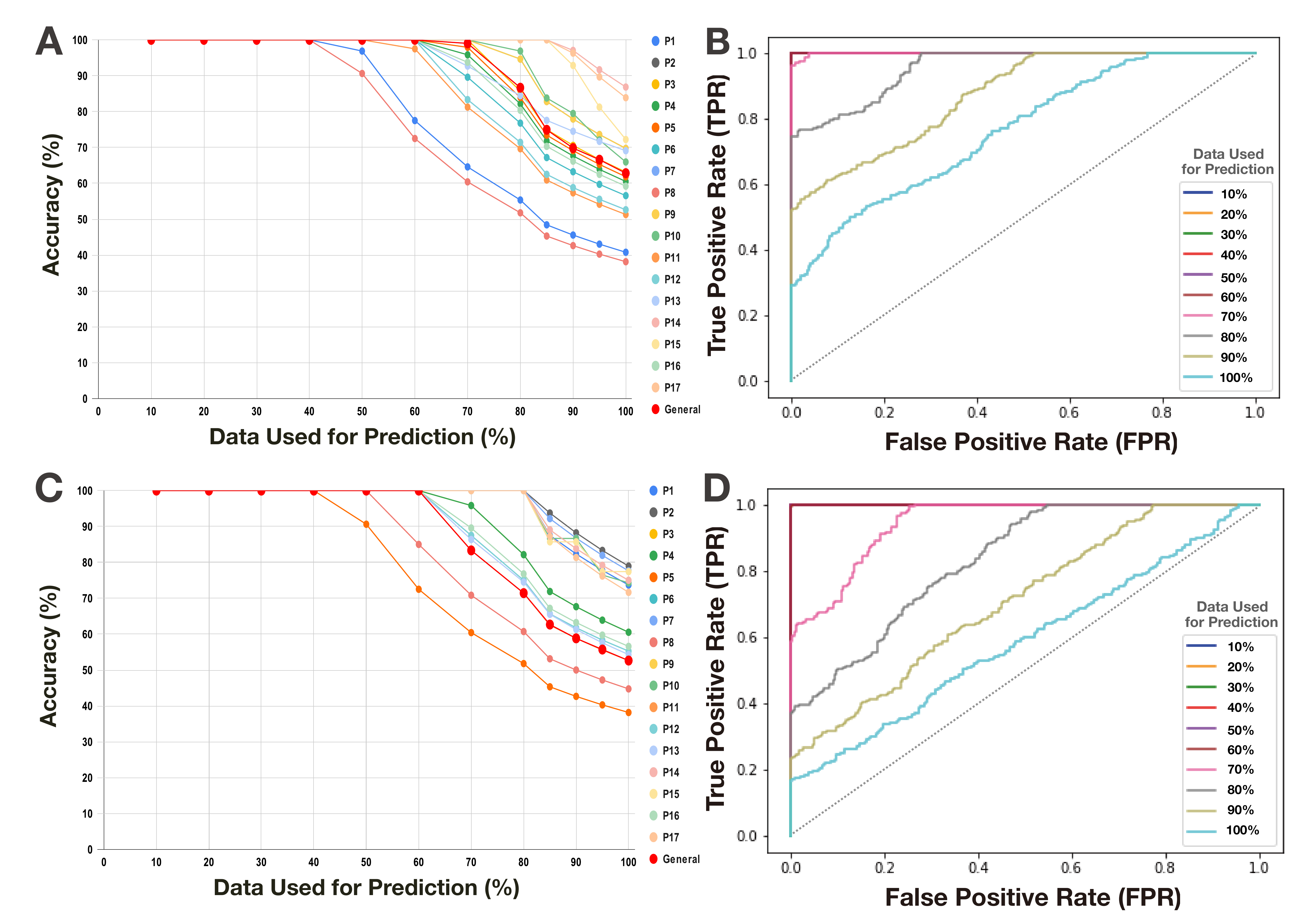}
\caption{
Accuracy and ROC curves for different percentages of data used for prediction from Annotation Task data (A and B) and Pointing Task data (C and D) in User Study. The red lines in (A) and (C) are the accuracy of the general models trained with all participants data, and the other lines are of the individual models. (B) and (D) are ROC curve of the general models for different percentage of data used according to their calculated confidence scores (\it PAT).}
\label{model_mobile}
\end{figure}

\begin{table}[b]
\caption{Weights for each smartphone feature.}
\small
\begin{tabular}{lllll}
\hline
\multirow{2}{*}{\textbf{Task}}       & \multirow{2}{*}{\textbf{No.}} & \multirow{2}{*}{\textbf{Features}}                           & \multicolumn{2}{l}{\textbf{Weight of the model}} \\ \cline{4-5} 
                            &                      &                                                     & \textbf{General}                       & \textbf{Within-user}                       \\ \hline
\multirow{2}{*}{Annotation} & 1                    & Tapping completion time from touch-down to touch-up & 0.03793                       & 0.06854                           \\
                            & 2                    & Maximum contact size                                & 0.07707                       & 0.000                             \\ \hline
\multirow{2}{*}{Pointing}   & 1                    & Tapping completion time from touch-down to touch-up & 0.4985                        & -0.2510                           \\
                            & 2                    & Maximum contact size                                & 0.1177                        & 0.6231                            \\ \hline
\end{tabular}
\label{table:weights_smartphone}
\end{table}

\begin{table}[b]
\caption{\markupchi{Average completion time (sec) for each task. The figures in the brackets mean {\it SD} of each average. PredicTaps shortens the time in \textit{Single Tap} cases, and it did not significantly increase the time in \textit{Double Tap} cases where false positives could occur.}}
\small
\begin{tabular}{lllll}
\hline
\multicolumn{1}{c}{\multirow{2}{*}{\textbf{PredicTaps}}} & \multicolumn{2}{c}{\textbf{Single Tap}}                                & \multicolumn{2}{c}{\textbf{Double Tap}}                                \\ \cline{2-5} 
\multicolumn{1}{c}{}                            & \multicolumn{1}{c}{\textbf{Annotation}} & \multicolumn{1}{c}{\textbf{Pointing}} & \multicolumn{1}{c}{\textbf{Annotation}} & \multicolumn{1}{c}{\textbf{Pointing}} \\ \hline
on                                              & 51.92 (8.459)                   & {\bf 24.89 (3.463)}                 & 56.78 (12.81)                   & 28.39 (7.377)                 \\
off                                             & 53.38 (10.38)                   & 29.24 (4.654)                 & 55.48 (9.345)                   & 26.58 (11.90)                 \\ \hline
\end{tabular}
\label{table:completiontime}
\end{table}

\subsection{Results}

\begin{table}[]
\caption{Results of Mann-Whitney U test on Accuracy on {\it Annotation Task} between groups divided by age or gender in the user study and average accuracies on the groups.}
\tiny
\begin{tabular}{lllllllllllll}
\hline
\multicolumn{3}{l}{\textbf{Data Used for Prediction (\%)}}                                                                                                                 & 10    & 20    & 30    & 40     & 50     & 60     & 70     & 80      & 90               & 100     \\ \hline
\multirow{2}{*}{\textbf{\begin{tabular}[c]{@{}l@{}}P-value of U test \\ Between men and women\end{tabular}}}     & \multicolumn{2}{l}{\textbf{Annotation}}                 & 1.00 & 1.00 & 1.00 & 0.571 & 0.901 & 0.937 & 0.676 & 0.343  & 0.326           & 0.367  \\ \cline{2-13} 
                                                                                                                 & \multicolumn{2}{l}{\textbf{Pointing}}                   & 1.00 & 1.00 & 1.00 & 1.00  & 0.351 & 0.742 & 0.809 & 0.810  & 0.808           & 0.810  \\ \hline
\multirow{4}{*}{\textbf{\begin{tabular}[c]{@{}l@{}}Average Accuracy\\ Between men and women\end{tabular}}}       & \multirow{2}{*}{\textbf{Annotation}} & \textbf{Men} & 100.0 & 100.0 & 100.0 & 99.60  & 97.18  & 91.01  & 83.99  & 78.63   & 69.73            & 61.77   \\ \cline{3-13} 
                                                                                                                 &                                      & \textbf{Women}   & 100.0 & 100.0 & 100.0 & 98.95  & 96.66  & 90.50  & 80.80  & 70.59   & 62.48            & 56.05   \\ \cline{2-13} 
                                                                                                                 & \multirow{2}{*}{\textbf{Pointing}}   & \textbf{Men} & 100.0 & 100.0 & 100.0 & 100.0  & 98.12  & 92.96  & 86.83  & 76.65   & 68.06            & 61.94   \\ \cline{3-13} 
                                                                                                                 &                                      & \textbf{Women}   & 100.0 & 100.0 & 100.0 & 98.95  & 96.94  & 90.38  & 83.87  & 74.06   & 65.74            & 59.87   \\ \hline
\multirow{2}{*}{\textbf{\begin{tabular}[c]{@{}l@{}}P-value of U test \\ Between older and younger\end{tabular}}} & \multicolumn{2}{l}{\textbf{Annotation}}                 & 1.00 & 1.00 & 1.00 & 0.227 & 0.191 & 0.164 & 0.123 & 0.0573 & \textbf{0.0488} & 0.0524 \\ \cline{2-13} 
                                                                                                                 & \multicolumn{2}{l}{\textbf{Pointing}}                   & 1.00 & 1.00 & 1.00 & 0.347 & 0.747 & 0.655 & 0.714 & 0.728  & 0.726           & 0.745  \\ \hline
\multirow{4}{*}{\textbf{\begin{tabular}[c]{@{}l@{}}Average Accuracy\\ Between older and younger\end{tabular}}}   & \multirow{2}{*}{\textbf{Annotation}} & \textbf{Younger} & 100.0 & 100.0 & 100.0 & 98.43  & 93.75  & 85.93  & 76.17  & 66.52   & {\bf 58.83}            & 52.75   \\ \cline{3-13} 
                                                                                                                 &                                      & \textbf{Older}   & 100.0 & 100.0 & 100.0 & 100.0  & 99.72  & 95.02  & 87.75  & 81.35   & {\bf 72.17}            & 64.07   \\ \cline{2-13} 
                                                                                                                 & \multirow{2}{*}{\textbf{Pointing}}   & \textbf{Younger} & 100.0 & 100.0 & 100.0 & 98.82  & 96.56  & 90.22  & 84.53  & 73.95   & 65.63            & 59.85   \\ \cline{3-13} 
                                                                                                                 &                                      & \textbf{Older}   & 100.0 & 100.0 & 100.0 & 100.0  & 98.33  & 92.82  & 85.91  & 76.46   & 67.91            & 61.72   \\ \hline
\end{tabular}
\label{table:pval_lab}
\end{table}

\begin{table}[]
\caption{Results of Mann-Whitney U test and Cohen's {\it d} between single- or double taps on {\it Annotation Task} divided by feature and age in the user study. The parentheses next to effect sizes indicate descriptors for magnitudes of {\it d}~\cite{cohen2013statistical, Sawilowsky2009}}
\small
\begin{tabular}{llll}
\hline
\textbf{Feature}                         & \textbf{Group} & \textbf{P-value}  & \textbf{Effect Size {\it d}} \\ \hline
\multirow{2}{*}{Tapping Completion Time} & Younger        & \textbf{0.00193}                        & 0.110 (Very Small)               \\ \cline{2-4} 
                                         & Older          & \textbf{4.24×10\textasciicircum{}(-57)} & 0.607 (Medium)                   \\ \hline
\multirow{2}{*}{Maximum Contact Size}    & Younger        & \textbf{4.36×10\textasciicircum{}(-8)}  & 0.195 (Very Small)               \\ \cline{2-4} 
                                         & Older          & 0.137                                   & 0.0441 (Very Small)              \\ \hline
\end{tabular}
\label{table:pval_feature_lab}
\end{table}

\begin{figure}[b!]
\centering
\includegraphics[width=0.8\columnwidth]{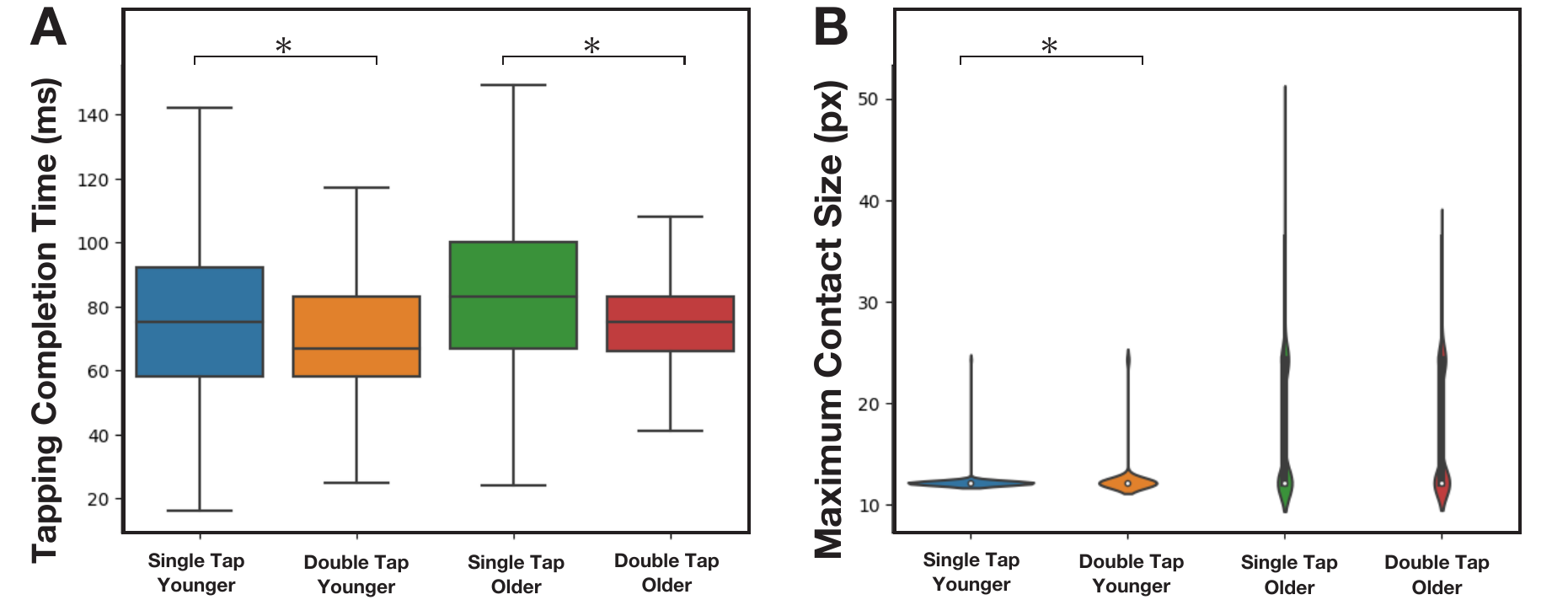}
\caption{
Box plot and violin plot of the tapping completion time and maximum contact size in {\it Annotation Task} (*: p<0.05) in the user study. A) Box plot of the tapping completion time on single- and double-tap divided by age. B) Violin plot of the maximum contact size on single- and double-tap divided by age.}
\label{fig:young_old_feature_labstudy}
\end{figure}

\subsubsection{Model Performance}
\label{subsubsec:smartphone_performance}
Figure \ref{model_mobile}(A) and (B) show the prediction accuracy using the data from the \textit{Annotation Task} and \textit{Pointing Task}, respectively. 
The red line indicates the general accuracy, and the other lines are the within-user accuracy. 
We again applied the \textit{PAT} to limit the data to high confidence scores. 
The accuracy here improved more rapidly than with the models of data from the laptop touchpad: it reached 100\,\% for every within-user case when we limited the data to the top 60\%, and 100\,\% in general when we limited the data to the top 50\,\%, as did the Precision and Recall.
Moreover, in the within-user case, all individual models could achieve more than 98.75\% accuracy with its top 80\% of data. 
Table~\ref{table:weights_smartphone} lists the features and their weights on the general models and two samples of the within-user models.
We found that the features varied among individuals and that the completion time and the maximum contact size had significant variations. Also, the general model's maximum contact size tendencies were opposite to those of the touchpad model discussed in Sec.\ref{sec:in-the-wild}, although the completion time tendencies were similar to the model with a longer completion time that is likely to be judged as a single tap.

As for the completion time for each task, Table~\ref{table:completiontime} shows the results of the average completion time for the {\it Annotation Task} and {\it Pointing Task} along with the {\it SD}. 
\markupcameraready{We observe average latency reductions of 1.5\,s on {\it Annotation Task} and 4.3\,s on {\it Pointing Task} per 20 taps. }
Between the on and off conditions of PredicTaps, only the {\it ( Single Tap / Pointing Task )} condition exhibited a significant difference (p = 0.0314 in Mann-Whitney's U test, effect size 1.05 in Cohen's d). 
\markupcameraready{The ideal latency reduction is $(500\,ms - 17.6\,ms)\times20\,taps = 9648\,ms \approx 9.6\,s$. Therefore, these results were reasonable, considering that most of the within-user models recorded roughly around 70 \% accuracy. In addition, the latency is more reduced in {\it Pointing Task} than in {\it Annotation Task} due to less cognitive noise, which is often biased by individual knowledge.
The double tap conditions show a slight increase in the completion time due to the false-positive cases, but no significant difference.}

\markupmobile{
We also analyzed the differences stemming from age and sex on tap features.
For this analysis, we divided the data into two groups (age: above and below median, sex: male and female) and then conducted the Mann-Whitney U test on the group, as none of the groups showed normality from the Shapiro-Wilk test. 
Table \ref{table:pval_lab} shows the results of the U tests on Accuracy divided by gender or age, and the average accuracy in those groups. 
Most of the accuracy did not show any difference (p>.05), but the p-values of the U test between elderly and young people were relatively low, and one of them showed a significant difference. The older group in {\it Annotation Task} had a tendency to have high accuracy.
Therefore, we looked into the features of older and younger groups.
Table~\ref{table:pval_feature_lab} shows the result of the Mann-Whitney U test and Cohen's d between single- and double-tap divided by feature and age. The bold p-values indicate that the features in the age group make a significant difference. 
Figure~\ref{fig:young_old_feature_labstudy} is the plot of the Table~\ref{table:pval_feature_lab}.
Here, we can deduce that Tapping Completion Time in the older group is a strong clue to predict whether a specific tap is a single- or a double-tap.
Therefore, we concluded that Tapping Completion Time makes a lot of difference between single tap and double tap among older people, although the individual difference is a more important factor to tap features.
}

\subsubsection{Usability Evaluation}
\label{subsubsec:smartphone_userstudy}
Table \ref{table:questionnaire_smartphone2} summarizes the questionnaire results. 
The latency reduction for single taps occurred for all participants. 
According to Q1 to Q3, 14 participants answered correctly for both {\it Annotation Task} and {\it Pointing Task}, and the results of Q2 and Q3 mean that they are convinced that PredicTaps reduces single tap latency. Q7 and Q8 suggest that 13 participants noticed the inconsistency of latency reduction of PredicTaps while single tapping, but they rather did not get confused by the inconsistency.
As for the effect of false-positive, the result of Q4 indicates that eight participants encountered the false-positive case in {\it Annotation Task} and 13 participants did in {\it Pointing Task}, though the result of Q5 indicates it's rather acceptable. 
Lastly, the result of Q6 indicates that the participants rated PredicTaps on smartphones as good.
\markupcameraready{To sum up, most participants noticed the latency reduction when single tapping and false-positives when double tapping, but accepted the false-positives (as seen in Q5). The overall score in Q6 was 5.588 ({\it SD}=0.8166) out of a 7-point Likert Scale, indicating a positive rating of PredicTaps. In Q7 and Q8, most participants noticed the inconsistency of the latency reduction but found it acceptable.}

\begin{figure}
\centering
\includegraphics[width=\columnwidth]{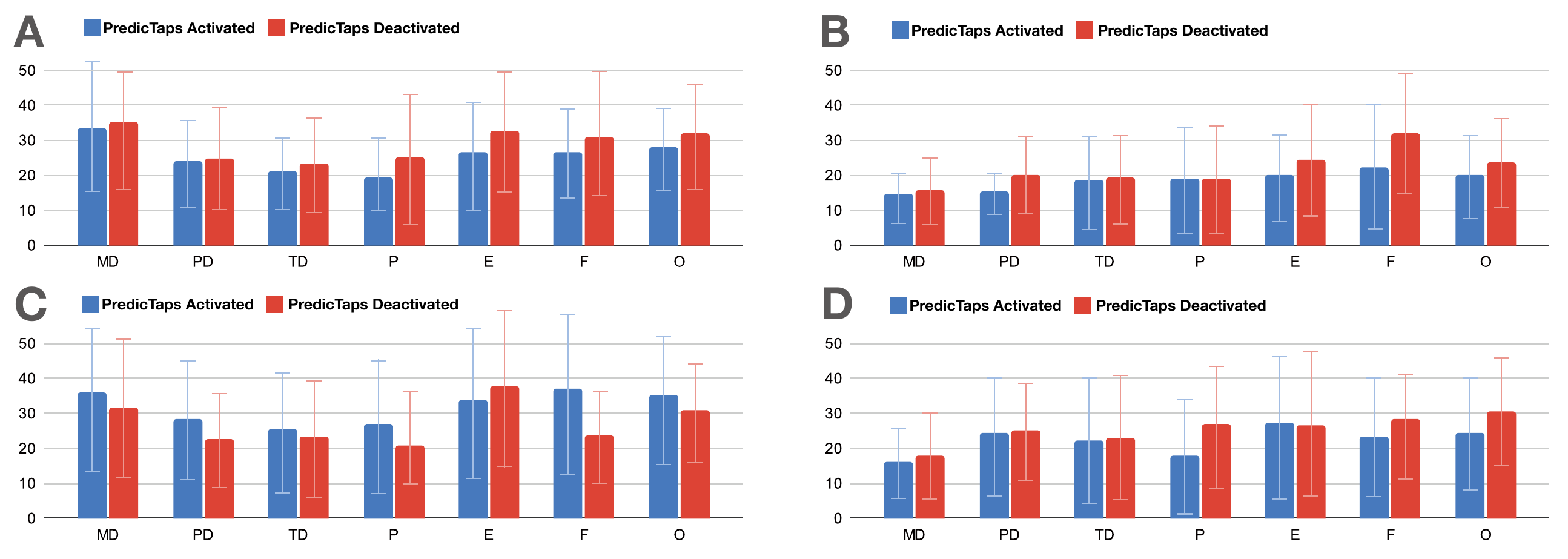}
\caption{
Results of NASA-TLX. MD means {\it Mental Demand}, PD means {\it Physical Demand}, TD means {\it Temporal Demand}, E means {\it Effort}, F means {\it Frustration Level}, and O means {\it Overall Score}. A){\it (Single Tap/Annotation Task)}. B){\it (Single Tap/Pointing Task)}. C){\it (Double Tap/Annotation Task)}. D){\it (Double Tap/Pointing Task)}. Error bars indicate SDs.}
\label{nasa-tlx}
\end{figure}

Fig. \ref{nasa-tlx} shows the result of NASA--TLX of each task by single- or double-tap. All of the measures mark no significant difference between where PredicTaps is activated and where it's not \markupmobile{(p>0.1 for all items). However, a slight decrease occurs in Effort and Frustration in {\it Single Tap} groups under the PredicTaps-activated condition. \markupcameraready{This is probably a result of PredicTaps' latency reduction.} Furthermore, in {\it ( Double Tap / Annotation Task )} groups, an increase can be seen in frustration under the PredicTaps-activated condition; \markupcameraready{possibly due to the false-positive cases.} In {\it ( Double Tap / Pointing Task )} groups, however, the frustration score decreases in the PredicTaps-activated condition.\markupcameraready{ We assume this is partly because the task feature, as {\it Pointing Task} does not require consideration and concentration. Therefore, they were not stressed that much.}}

\markupcameraready{In the post-experiment interview, some participants mentioned annoyance with image loading in {\it Annotation Task} when PredicTaps fastened the reaction time. As for this opinion, PredicTaps' prediction can be used to enhance throughput or reduction of reaction time, such as process scheduling as future work.
Another participant also mentioned the variation in accuracy depending on the tap location. 
In the smartphone model, we did not include features that could invade privacy, such as using locations of tapping points as features. However, to achieve high accuracy and better usability, feature engineering for PredicTaps should be conducted in the future.}

\begin{table}[b]
\caption{\markupmobile{Number of correct and wrong replies to Q1 and Q4, as well average value of replies to Q2, Q3, Q5, and Q6. The figures in the brackets mean {\it SD} of each average.}}
\footnotesize
\begin{tabular}{llll}
\hline
\multirow{2}{*}{} & \multicolumn{1}{c}{\multirow{2}{*}{\textbf{Question}}}                                                                                                      & \multicolumn{2}{c}{\textbf{Reply}}                                               \\ \cline{3-4} 
                  & \multicolumn{1}{c}{}                                                                                                                               & \multicolumn{1}{c}{\textbf{Annotation Task}} & \multicolumn{1}{c}{\textbf{Pointing Task}} \\ \hline
Q1                & Which partial section did you think PredicTaps is activated?                                                                                      & 14 correct / 3 wrong                & 14 correct / 3 wrong              \\ \hline
Q2                & Rate your confidence level of the answer to Q1.                                                                                                    & 4.529 (1.550)                       & 5.471 (1.156)                     \\ \hline
Q3                & To what extent do you think the single-tap latency decreases?                                                                                      & 5.176 (1.051)                       & 5.529 (1.377)                     \\ \hline
Q4                & Were your double taps wrongly recognized by the system?                                                                                           & 8 Yes / 9 No                        & 13 Yes / 4 No                     \\ \hline
Q5                & (If "yes" to Q4) To what extent can you accept the misjudgment?                                                                                   & 5.467(1.433)                        & 4.647 (1.550)                     \\ \hline
Q6                & Rate the usefulness of PredicTaps.                                                                                                                 & \multicolumn{2}{c}{5.588 (0.8166)}                                      \\ \hline
Q7                & \begin{tabular}[c]{@{}l@{}}When single tapping, did you feel the difference \\ between when PredicTaps activated and when it did not?\end{tabular} & \multicolumn{2}{c}{13 Yes / 4 No}                                       \\ \hline
Q8                & (If "yes" to Q7) Did you feel confused about the inconsistency?                                                                            & \multicolumn{2}{c}{2.929 (1.685)}                                       \\ \hline
\end{tabular}
\label{table:questionnaire_smartphone2}
\end{table}

\section{PredicTaps on Daily Smartphone Use}
\label{sec:smartphone_daily}

In addition to the user experience of PredicTaps, in order to evaluate the robustness of PredicTaps by form factors, we conducted an experiment to assess PredicTaps performance on smartphones under various conditions.

\subsection{Participants and Apparatus}
\markupmobile{
We recruited 17 participants from a broad range of ages and sex (seven women, nine men, 14 right-handed, two left-handed, and one ambidextrous, average age 34.12 with {\it SD} = 16.31) and let them use their own smartphones. 
Our objectives here were to ensure the generalizability of the system and to investigate age-specific or sex-specific features. 
The participants used iPhone SE, iPhone 8, iPhone XR, iPhone12 mini, iPhone pro max 12, iPhone 13 Pro, and iPad, running iOS 15--16. 
The average number of years the participants had used smartphones was 8.562 with {\it SD} = 2.312). 
The average number of hours the participants used smartphones per day was 5.500 with {\it SD} = 2.875). 
We informed the participants that this experiment was to log the operation of the smartphone and obtained their consent to participate. 
}

\subsection{Task Procedure and Data Collection}
\markupmobile{
Although it is essential to investigate the performance of PredicTaps on any occasion, it is difficult to have smartphone users utilize daemon apps to retrieve daily tap data because it can invade their sense of privacy and security. 
For example, we might be able to calculate passwords or text messages that users type from the locations of tap data. 
Therefore, we designed the setting of this data collection to be as close to an in-the-wild condition as possible. 
To assess the robustness of PredicTaps under in-the-wild conditions, we changed the following conditions from the usability experiment in the lab:
\begin{itemize}
    \item We did not limit the participants' postures, such as the hand holding the smartphone, or the finger tapping the screen. 
    \item The participants performed tasks randomly during the daytime for various experimental occasions (e.g., while walking, sitting, and in cars).
\end{itemize}
}


We developed an iOS app to encourage participants to perform the task applications (\textit{Annotation Task} / \textit{Pointing Task})\markupcameraready{, which sends users notifications via Lock Screen, Notification Center, and Banners.} 
We used the same task applications as in the user study in Sec.~\ref{sec:smartphone} (Fig. \ref{ui_mobile}). 
Participants installed the app and ran it while performing daily work on their smartphones for six days. 
The app prompted the participants to perform the task applications every hour, and the participants were required to perform the task applications five times a day. 
In each game, participants were required to tap ten times. 
The threshold to distinguish a single tap from a double tap was set to 500 ms. 
We collected data for a total of 600 taps (single tap: ten taps for each participant $\times$ five games $\times$ six days; double tap: ten taps for each participant $\times$ five games $\times$ six days). 
None of the participants used triple taps for operations.

\begin{figure}[b!]
\centering
\includegraphics[width=\columnwidth]{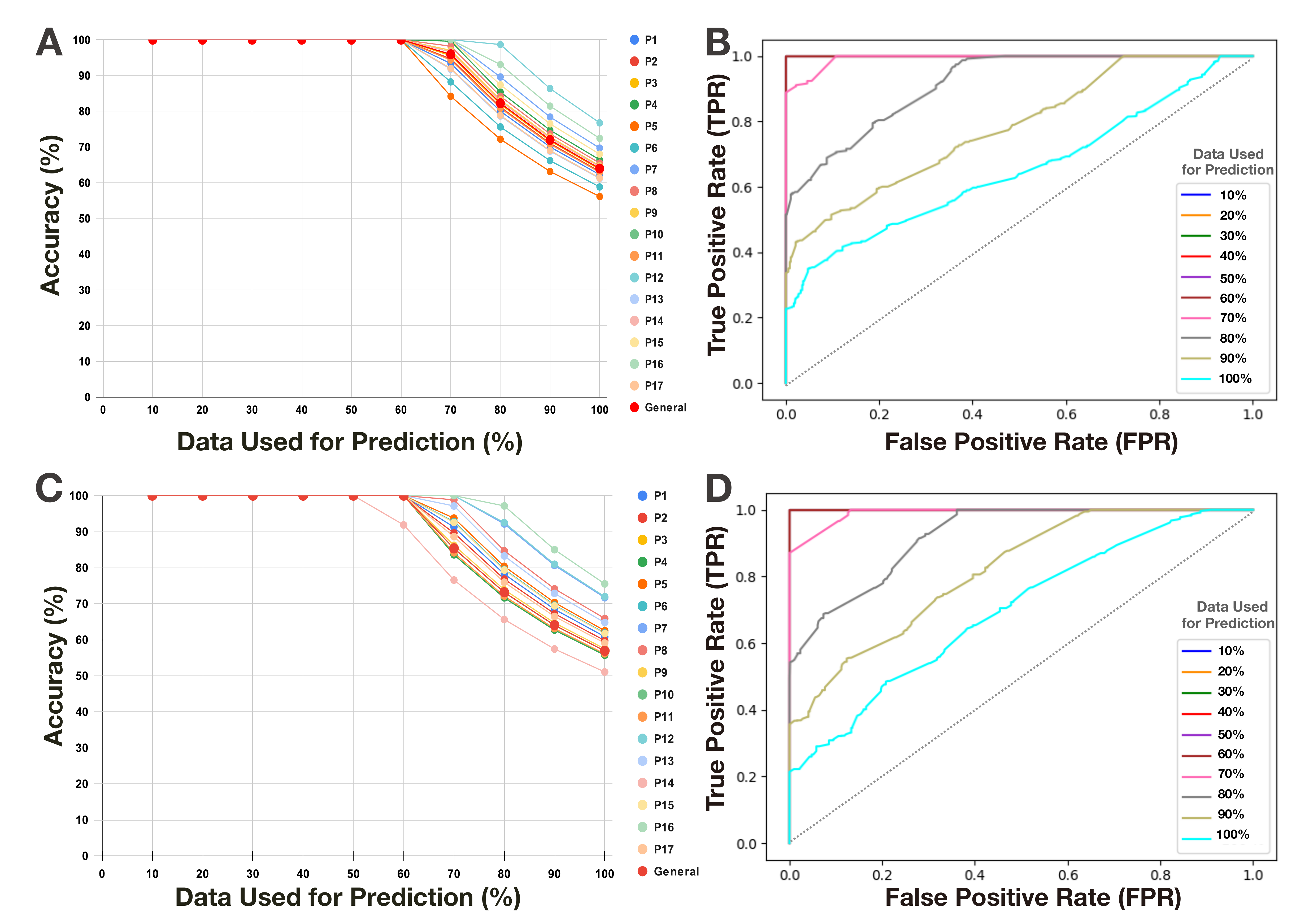}
\caption{
Accuracy and ROC curves for different percentages of data used for prediction from Annotation Task data (A and B) and Pointing Task data (C and D) in the Daily Smartphone Experiment. The red lines in (A) and (C) are the accuracy of the general models trained with all participants' data, and the other lines are of the individual models. (B) and (D) are ROC curve of the general models for a different percentage of data used according to their calculated confidence scores (\it PAT).}
\label{model_mobile_inthewild}
\end{figure}

\begin{table}[]
\caption{Results of Mann-Whitney U test on Accuracy between groups divided by age or gender in the in-the-wild experiment.}
\tiny
\begin{tabular}{llllllrrrrrrr}
\hline
\multicolumn{3}{l}{\textbf{Data Used for Prediction (\%)}} &
  10 &
  20 &
  30 &
  \multicolumn{1}{l}{40} &
  \multicolumn{1}{l}{50} &
  \multicolumn{1}{l}{60} &
  \multicolumn{1}{l}{70} &
  \multicolumn{1}{l}{80} &
  \multicolumn{1}{l}{90} &
  \multicolumn{1}{l}{100} \\ \hline
\multirow{2}{*}{\textbf{\begin{tabular}[c]{@{}l@{}}P-value of U test \\ Between men and women\end{tabular}}} &
  \multicolumn{2}{l}{\textbf{Annotation}} &
  1.00 &
  1.00 &
  1.00 &
  \multicolumn{1}{l}{1.00} &
  1.00 &
  0.225 &
  0.320 &
  0.322 &
  0.322 &
  0.345 \\ \cline{2-13} 
 &
  \multicolumn{2}{l}{\textbf{Pointing}} &
  1.00 &
  1.00 &
  1.00 &
  \multicolumn{1}{l}{1.00} &
  0.350 &
  0.742 &
  0.810 &
  0.809 &
  0.807 &
  0.809 \\ \hline
\multirow{4}{*}{\textbf{\begin{tabular}[c]{@{}l@{}}Average Accuracy\\ Between men and women\end{tabular}}} &
  \multirow{2}{*}{\textbf{Annotation}} &
  \textbf{Men} &
  100.0 &
  100.0 &
  100.0 &
  100 &
  100 &
  96.48 &
  84.56 &
  73.99 &
  65.77 &
  59.19 \\ \cline{3-13} 
 &
   &
  \textbf{Women} &
  100.0 &
  100.0 &
  100.0 &
  100 &
  100 &
  93.61 &
  81.36 &
  71.19 &
  63.28 &
  56.95 \\ \cline{2-13} 
 &
  \multirow{2}{*}{\textbf{Pointing}} &
  \textbf{Men} &
  100.0 &
  100.0 &
  \multicolumn{1}{r}{100} &
  100 &
  100 &
  100 &
  91.39 &
  79.80 &
  69.83 &
  62.07 \\ \cline{3-13} 
 &
   &
  \textbf{Women} &
  100.0 &
  100.0 &
  \multicolumn{1}{r}{100} &
  100 &
  100 &
  98.98 &
  90.22 &
  78.76 &
  68.92 &
  61.25 \\ \hline
\multirow{2}{*}{\textbf{\begin{tabular}[c]{@{}l@{}}P-value of U test \\ Between older and younger\end{tabular}}} &
  \multicolumn{2}{l}{\textbf{Annotation}} &
  1.00 &
  1.00 &
  1.00 &
  \multicolumn{1}{l}{1.00} &
  1.00 &
  0.119 &
  0.0538 &
  0.0540 &
  0.0524 &
  0.0538 \\ \cline{2-13} 
 &
  \multicolumn{2}{l}{\textbf{Pointing}} &
  1.00 &
  1.00 &
  1.00 &
  1.00 &
  0.347 &
  0.681 &
  0.336 &
  0.336 &
  0.335 &
  0.333 \\ \hline
\multirow{4}{*}{\textbf{\begin{tabular}[c]{@{}l@{}}Average Accuracy\\ Between older and younger\end{tabular}}} &
  \multirow{2}{*}{\textbf{Annotation}} &
  \textbf{Younger} &
  100.0 &
  100.0 &
  \multicolumn{1}{r}{100.0} &
  100.0 &
  100.0 &
  93.28 &
  79.96 &
  69.96 &
  62.19 &
  55.97 \\ \cline{3-13} 
 &
   &
  \textbf{Older} &
  100.0 &
  100.0 &
  100.0 &
  100.0 &
  100.0 &
  96.77 &
  85.81 &
  75.09 &
  66.74 &
  60.07 \\ \cline{2-13} 
 &
  \multirow{2}{*}{\textbf{Pointing}} &
  \textbf{Younger} &
  100.0 &
  100.0 &
  100.0 &
  100.0 &
  100.0 &
  90.08 &
  77.21 &
  67.56 &
  60.05 &
  54.05 \\ \cline{3-13} 
 &
   &
  \textbf{Older} &
  100.0 &
  100.0 &
  100.0 &
  100.0 &
  99.09 &
  91.52 &
  81.18 &
  71.03 &
  63.13 &
  56.82 \\ \hline
\end{tabular}
\label{table:pval_daily}
\end{table}

\begin{table}[]
\caption{Results of Mann-Whitney U test and Cohen's {\it d} between single- or double taps on {\it Annotation Task} divided by feature and age in data collection under the in-the-wild condition. The parentheses next to effect sizes indicate descriptors for magnitudes of {\it d}~\cite{cohen2013statistical, Sawilowsky2009}}
\small
\begin{tabular}{llll}
\hline
\textbf{Feature}                      & \textbf{Group} & \textbf{P-value in Mann Whitney U test}   & \textbf{Effect Size in Cohen's d} \\ \hline
\multirow{2}{*}{Tapping Completion Time} & Younger & \textbf{0.00}                            & 0.294 (Small)      \\ \cline{2-4} 
                                         & Older   & \textbf{0.00}                            & 0.875 (Large)      \\ \hline
\multirow{2}{*}{Maximum Contact Size} & Younger        & \textbf{7.09×10\textasciicircum{}(-180)} & 0.114 (Very Small)               \\ \cline{2-4} 
                                         & Older   & \textbf{8.93×10\textasciicircum{}(-250)} & 0.115 (Very Small) \\ \hline
\end{tabular}
\label{table:pval_feature_inthewild}
\end{table}

\begin{figure}[t!]
\centering
\includegraphics[width=0.8\columnwidth]{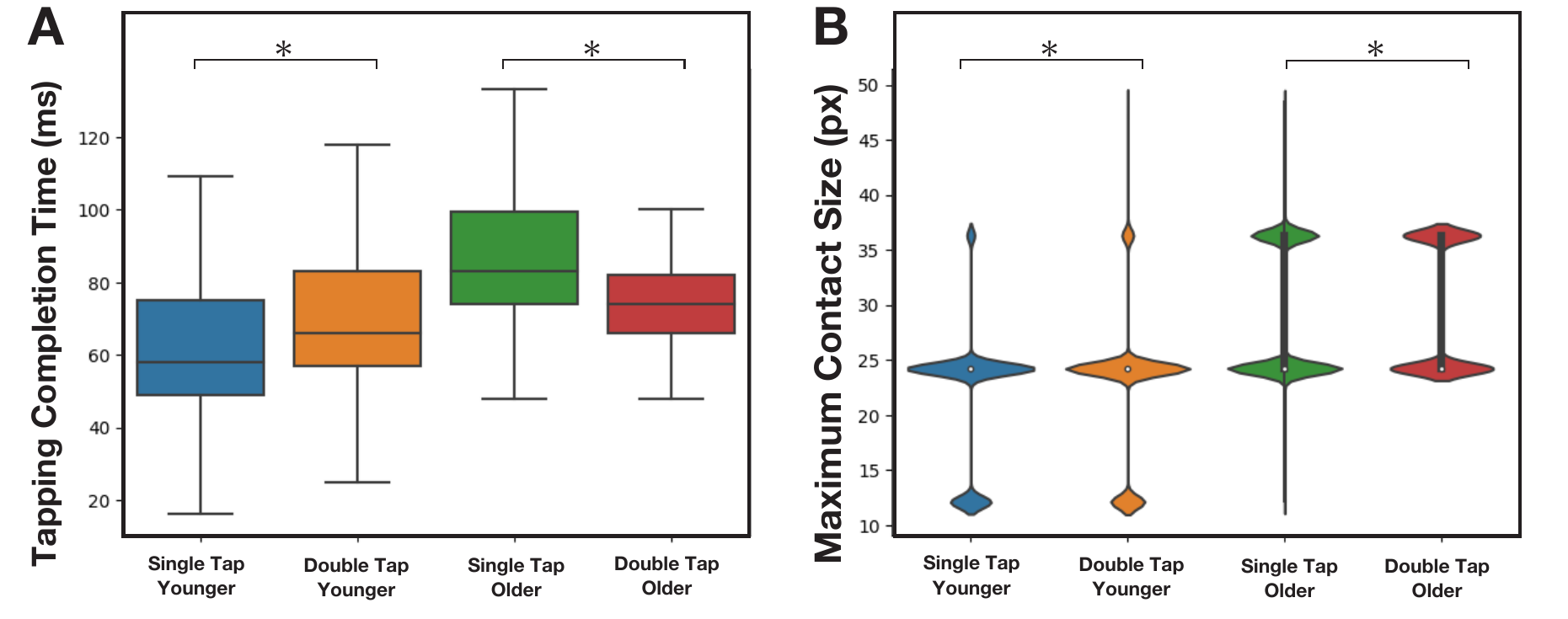}
\caption{
Box plot and violin plot of the tapping completion time and maximum contact size in {\it Annotation Task} (*: p<0.05) in data collection under the in-the-wild condition. A) Box plot of the tapping completion time on single- and double-tap divided by age. B) Violin plot of the maximum contact size on single- and double-tap divided by age.}
\label{fig:young_old_feature_inthewild}
\end{figure}

\subsection{Results}
\label{sec:smartphone_inthewild_result}
\markupmobile{We conducted the data processing in the same manner as discussed in Sec.~\ref{sec:learn} and Sec.~\ref{sec:in-the-wild}.
Figure \ref{model_mobile_inthewild}(A) and (B) show the prediction accuracy using the data from the \textit{Annotation Task} and from the \textit{Pointing Task}, as in Sec~\ref{sec:smartphone}. 
}
\markupmobile{The within-user accuracy and the general accuracy are as same as described in Sec.~\ref{sec:in-the-wild}. 
The {\it PAT} was applied to limit the data being used to high scores. 
Same as in Sec.~\ref{sec:in-the-wild} and Sec.~\ref{sec:smartphone}, the accuracy gradually improved, reaching 95\% for the average of within-user cases when we limited the data to the top 50\% to 60\%, though the accuracy is worse than the previous models in Sec.~\ref{sec:in-the-wild} and Sec.~\ref{sec:smartphone}. 
}

\markupmobile{
We also analyzed the difference coming from age and sex on tap features.
Therefore, we divided the data into two groups (age: age above median and age below median, sex: men and women). Then, we conducted the Mann-Whitney U test on the group because none of the groups showed normality from the Kolmogorov–Smirnov test.
Table \ref{table:pval_daily} shows the results of the U tests on Accuracy divided by gender or age and the average accuracies of each group.
All accuracy did not show differences (p>0.05). However, the p-values of the U test between elderly and young people were relatively low, and one showed a significant difference. 
Therefore, we looked into the features of elderly people and young people like in the user study in Sec.~\ref{subsubsec:smartphone_performance}.
Therefore, we concluded that individual difference is a more important factor in tapping features.
Table~\ref{table:pval_feature_inthewild} shows the result of the Mann-Whitney U test and Cohen's d between single- and double-tap divided by feature and age. 
Figure~\ref{fig:young_old_feature_inthewild} is the plot of the Table~\ref{table:pval_feature_inthewild}.
Like in Sec.~\ref{subsubsec:smartphone_performance}, we can deduce that Tapping Completion Time in the older group is a strong clue to predict whether a specific tap is a single- or a double-tap.
Therefore, we came to a conclusion same as in Sec.~\ref{subsubsec:smartphone_performance}; Tapping Completion Time makes a lot of difference between single tap and double tap among older people, although the individual difference is a more important factor to tap features.
}

\section{Discussion}
\subsection{PredicTaps' Applicability for Different Data Conditions and Form Factors}

We conducted three experiments on touchpad and smartphone, including daily laptop use (Sec. \ref{sec:in-the-wild}), the \textit{Annotation Task}, and the \textit{Pointing Task}.
Despite our assumption that a model's performance would be worse if it is trained with quick tapping data because there would be less difference between single- and double-taps in terms of the completion time of the tap and maximum tap size. However, all models recognize the difference between single- and double-taps under varying tap frequencies (low: \textit{Annotation Task}, high: \textit{Pointing Task}). 

Moreover, PredicTaps performed well on the touchpad and smartphone experiments, showing applicability to different form factors. The basic features for training were consistent, while other features varied depending on the hardware of the touch surfaces (e.g., AC adopter or battery condition on laptop touchpad).


\subsection{PredicTaps Models' Performance}

\markupcameraready{As for smartphone models,} We found that the performance of PredicTaps was not significantly different compared to touchpads. However, smartphones have a greater variety of sensors than laptops (e.g., nine\,DoF sensors). 
In fact, we engineered this feature as well, but since we extracted only 400 taps per user, the overfitting of the models occurred, leading to bad usability overall. 
Therefore, further evaluations are necessary for smartphone models to examine conditions such as participants walking, in a car, or tapping their smartphones with their thumb, to extract non-biased nine DoF sensor data. 

Also, we could potentially improve the \markupcameraready{smartphone} PredicTaps performance by utilizing a model that can process time-series data for processing nine\,DoF sensors of smartphones.
Combining it with a logistic regression model may improve the PredicTaps performance.

As for the general models \markupcameraready{on touchpad and smartphones}, they had poor accuracy on both touchpad and smartphone when using 100\,\% of the data and even had a worse {\it PAT}. However, this highlights the significant differences in individuals' tap features, which are potentially useful for authentication applications. 
Combined with recent bionic authentication methods (e.g., biosignals and irises), it may be possible to develop a solid new approach to user authentication~\cite{Zhang:2022:10.1145/3491102.3517440, Jiokeng:2022:10.1145/3534575}.

\markupmobile{In the daily smartphone experiment in Sec.~\ref{sec:smartphone_daily}, the accuracy was worse than the models \markupcameraready{of touchpad} in Secs.~\ref{sec:in-the-wild} and \markupcameraready{of smartphone under strict posture restrictions in the lab in Sec.}~\ref{sec:smartphone}. 
Therefore, we examined the difference between the lab study and in-the-wild conditions. 
We conducted the Mann-Whitney U test on the accuracy of each group according to {\it PAT} because none of the groups showed normality from the Shapiro-Wilk test. 
As for the individual models, which are the models trained with all the participants, there was no difference between the lab study and in-the-wild conditions (p>0.15). 
We conclude that this is because of the trade-off between the amount of training data and the flexibility of the posture of the fingers: in contrast to the lab experiment, we did not tell participants how to hold the smartphones, but we collected 1.5 times as many taps as in the in-the-wild condition.
}

\markupmobile{
As for age, we found the accuracy of models with older people is significantly better than models trained with young people's taps \markupcameraready{on smartphones} (Sec.~\ref{subsubsec:smartphone_performance} and Sec.~\ref{sec:smartphone_inthewild_result}).
Therefore, we came to the conclusion that older people tend to single tap significantly slower than the first tap of a double tap.
}

\markupmobile{
In the future, PredicTaps can also be finetuned based on the data of taps in users' everyday smartphone uses to improve its accuracy without much of any users burden. 
For example, PredicTaps could collect the tap data during users' smartphone uses in everyday lives, and have users record some tap data for a specific downstream task to finetune the model weights. This self-supervised method is already widely applied to gesture recognition and classification tasks in the field of Human-Computer Interaction~\cite{Su:2023:10.1145/3544548.3581465, Kimura:2022:10.1145/3526114.3558707, Wu:2020:10.1145/3313831.3376875, Gierad:2018:10.1145/3242587.3242609}. 
}

\markupchi{
\subsection{Amount of Training Data}
The amount of required training data is one of the problems of status quo PredicTaps.
During the data collection experiment on smartphones (Sec. \ref{sec:smartphone}), some participants mentioned the burden of tapping 400 times in total. Previous work has demonstrated that fatigue affects the time for tapping~\cite{Fatigue:10.1145/3313831.3376588}. 
In addition, the weights are very different in magnitude (e.g., 0.03793 vs. 0.4985), which indicates a possible dependence of the model on the application and the user, meaning that fine-tuning data should be collected from each user individually. 
Therefore, we should devise solutions to alleviate the user's workload when collecting the data. Implementing an interesting game application(e.g., the one used in ~\cite{Henze2011-rh}) would be one option to ease the mental burden on users. Also, we assume that few-shot learning can reduce the required training data: architectures that have fundamental models pretrained with massive data are put in place beforehand, and when it comes to actual implementation, they are fine-tuned with smaller in situ data to optimize for individuals~\cite{Wu:2020:10.1145/3313831.3376875, Xu:2022:10.1145/3491102.3501904, Kimura:10.1145/3526114.3558707}. 
In practical terms, integrating tabular data (e.g., the maximum contact size and the completion time) into a few-shot learning model would be one way to provide this kind of architecture.}
Moreover, if we input a short amount of user-specific tap data into fundamental models trained on a specific application (e.g., chat keyboard, clicking buttons), we can generate a user- and application-specific PredicTaps model that enables users to operate PredicTaps with less amount of data (i.e., fewer taps).

\markupchi{
\subsection{User Perception of Latency Inconsistency}
The smartphone user study in Sec.\ref{subsubsec:smartphone_userstudy} revealed that latency inconsistency confused some users, possibly due to model accuracy. We obtained a comment that ``Ironically, it sometimes felt like the system reaction slowed down when the system was not activated, and I noticed that in sections without PredicTaps it was not sometimes slow, rather it was fast.'' The overall evaluation of PredicTaps was positive, so the system's benefits outweigh the confusion. However, further evaluation is needed to assess the correlation between inconsistency and confusion, and its effect on usability.}

\subsection{Latency Reduction by PredicTaps}
We ran PredicTaps model on iPhone (13 and 11 pro max) and confirmed that data processing to determine single- or double-tap had an average latency of 1.38 ms ± 18.1 µs (mean ± std. dev. of 7 runs, average of 1000 trials).
Therefore, the reduced latency for a single tap can be calculated as a {\it double-tap threshold} (500 ms) -- processing time (1 ms for prediction and 11 ms for sensing on touchpad, 16.6 ms on smartphone; in total: 12 ms on touchpad and 17.6 ms on smartphone). 
This confirms the practicality of the logistic model used. More complex models may improve accuracy but at the cost of increased processing time.


\markupchi{In the user evaluation of smartphone PredicTaps, Latency reduction of 1.5\,s on {\it Annotation Task} and 4.3\,s on {\it Pointing Task} per 20 taps were observed.
Thus, we can infer that PredicTaps' small latency reduction in frequently used operations, such as single tap, is beneficial in the long run.}

\subsection{Implementation Layer}

\markupmobile{PredicTaps is a high-level optimization approach that operates distinctly from middleware-level optimizations, such as those associated with the kernel, driver, or operating system layers. 
The efficacy of PredicTaps as a higher-level implementation lies in its capacity to leverage information derived from high-level application programming interfaces (APIs) which have proven to be useful in previous works~\cite{Henze:2016:STL:2935334.2935381}. 
Consequently, this affords the advantage of being universally applicable across a diverse array of software systems, thereby enhancing its adaptability and overall utility within the realm of optimization strategies.}


\begin{table}[]
\caption{Results of Mann-Whitney U test and between the user study and the in-the-wild study on the accuracy on {\it Annotation Task} and on {\it Pointing Task}.}
\small
\begin{tabular}{llrrrrrrrrrr}
\hline
\multicolumn{2}{l}{\textbf{\begin{tabular}[c]{@{}l@{}}Data Used for \\ Prediction (\%)\end{tabular}}} &
  10 &
  20 &
  30 &
  40 &
  50 &
  60 &
  70 &
  80 &
  90 &
  100 \\ \hline
\multirow{2}{*}{\textbf{P-value}} &
  \textbf{Annotation} &
  1.00 &
  1.00 &
  1.00 &
  0.216 &
  0.145 &
  0.189 &
  0.850 &
  0.689 &
  0.727 &
  0.850 \\ \cline{2-12} 
 &
  \textbf{Pointing} &
  1.00 &
  1.00 &
  1.00 &
  0.332 &
  0.290 &
  0.819 &
  0.187 &
  0.154 &
  0.159 &
  0.128 \\ \hline
\end{tabular}
\label{tab:pvalue_userstudy_inthewild}
\end{table}

\subsection{The Difference of Model Performances between User Study and In-the-wild Study}
As for the difference of the model performance between the user study (Lab study) in Sec.~\ref{sec:smartphone} and in-the-wild study in Sec.~\ref{sec:smartphone_daily}, we conducted the Mann-Whitney U test on the accuracy between both studies according to {\it PAT} because none of the groups showed normality from the Korgomorov-Smirnov test. 
However, we could not found significant differences. Table \ref{tab:pvalue_userstudy_inthewild} shows the p-values of the Mann-Whitney U tests.

\section{Limitation}


\markupchi{As for the cost of false-positives, in some situations, the cost of the reverting operation is so massive (e.g., the buy confirmation button on EC sites) that it may negate the benefits, so the application of PredicTaps should be limited to situations where the reverting operation is easy to do. In the smartphone user study (Sec.\ref{day2_smartphone}), users could not proceed to the next task on a false positive case and had to perform the operation again. For such easy return operations, we proved that the benefits of reducing latency outweighed the disadvantages of false positives~(Sec.\ref{subsubsec:smartphone_userstudy}).}

\section{Conclusion and Future Directions}
In this paper, in order to solve the {\it single-tap latency problem}, we developed a method called PredicTaps and assess it in various conditions. The proposed method adopts a machine learning technique to distinguish a single tap from the first tap of a double tap by using the sensor data of touch surfaces. The model was trained from touch event--related data collected experimentally in three situations on two form factors (touchpad and smartphone).

We also found that the features could vary depending on the form factors, but even so, PredicTaps was effective on both a smartphone touch display and a laptop touchpad. It also succeeded in reducing the {\it single-tap latency} to 12 ms on the touchpad and to 17.6 ms on the smartphone, though the latency was typically about 150--500 ms. 
In the user study, most of the participants noticed the latency reduction enabled by our proposed method (13 out of 17 participants), and 14 out of 17 participants positively rated PredicTaps (responded with 5--7 for Q. 6 in the user study).


\bibliographystyle{ACM-Reference-Format}
\bibliography{acmart}

\newpage
\appendix
\section{Extraction of prediction results with top n \% of the score}
\label{sec:appendix}
\begin{algorithm}[]
\small
\label{alg-predictaps}
\begin{algorithmic}[1]
\Require{\\$array$: array that contains confidence scores according to each prediction for taps and the tap's data\\ $n$: threshold of the data extraction}
\Ensure{\\$top\_n\_array$: extracted data with top n \% of the scores}

\Function {calculate\_distance}{$array[[s_0, data_0], [s_1,data_1], ... [s_t,data_t], ... [s_n, data_n]]$}
    \ForAll {$e, i \gets array$} 
        \State $e \gets [|s_i - 0.5|, data_i]$
    \EndFor
    \State \Return $array$
\EndFunction

\Function {extract\_top\_n\_percent\_data}{$array[[s_0, data_0], [s_1,data_1], ... [s_t,data_t], ... [s_n, data_n]], n$}
    \State $array \gets \Call{calculate\_distance}{array} $
    \State $ array \gets \Call{sort}{array}$
    \State $top\_n\_array = array[0, n/100+1]$
    \State \Return $top\_n\_array$
\EndFunction

\end{algorithmic}
\end{algorithm}

\end{document}